\documentclass[%
 reprint,
showkeys,preprintnumbers,
nofootinbib,
nobibnotes,
 amsmath,amssymb,
 aps,
]{revtex4-1}
\usepackage{graphicx}
\usepackage{bm}
\usepackage{hyperref}
\usepackage[mathlines]{lineno}

\begin{document}

\preprint{APS/123-QED}

\title{Power spectrum of post inflationary  primordial magnetic fields}

\author{H\'ector J. Hort\'ua}
 \email{hjhortuao@unal.edu.co}
\author{Leonardo Casta\~neda}%
 \email{lcastanedac@unal.edu.co}
\affiliation{%
 Grupo de Gravitaci\'on y Cosmolog\'ia, Observatorio Astron\'omico Nacional,\\
 Universidad Nacional de Colombia, cra 45 $\#$26-85, Ed. Uriel Gutierr\'ez, Bogot\'a D.C, Colombia
}%



\begin{abstract}
The origin of large scale magnetic fields is one of the most puzzling topics in cosmology and astrophysics. 
It is assumed that the observed magnetic fields result from the amplification of an initial  field produced in the early universe.
In this paper we compute the exact power spectrum of  magnetic fields created after inflation  best known as post inflationary magnetic fields, using the first order cosmological perturbation theory. Our treatment differs from others works because we include an infrared cutoff which encodes only causal modes in the 
spectrum. The cross-correlation between magnetic energy density  with Lorentz force and the anisotropic part of the electromagnetic field are exactly computed.
We compare our results with previous works finding agreement in cases where the ratio between lower and upper cutoff is very small. However, we found that spectrum is strongly affected when this ratio is greather than 0.2. Moreover, the effect of a post inflationary magnetic field with a lower cutoff on the angular power spectrum in the temperature distribution of CMB was also exactly calculated.  The main feature is a shift of the  spectrum's peak  as  function of the infrared cutoff, therefore analyzing this effect we could infer the value of this cutoff and thus constraining the primordial magnetic fields generation models.
\begin{description}

\item[PACS numbers]
98.80.-k, 95.30.Qd.
98.80.Cq, 98.70.Vc, 98.80.-k.
\end{description}
\end{abstract}

\maketitle


\section{Introduction}
Magnetic fields have been observed in all scales of the universe, from  planets and stars to galaxies and galaxy clusters with strength of the
order of $10^{-6}$G at typical scales of $10$kpc \cite{Lawrence}. Also a lower
bound $3 \times 10^{-16}$G on the strength of magnetic fields in voids of the large scale
structure has been reported  from gamma-ray observations \cite{neronov}.  However, the origin of such a magnetic field remains as one of the  unsolved mysteries in modern cosmology. There is  a school of thought which states that magnetic fields we observe today have a primordial origin, indeed, there are some  processes in  early epoch of the universe that would have created a small magnetic field {\it a seed} and  after a while  possibly was amplified by dynamo actions or adiabatic compression during the structure formation era \cite{Kandus}, \cite{banerjee1}.  The evolution of this seed from its generation to the present has been   discussed in detail by \cite{barrow}, \cite{sigl2}, \cite{banerjee}, \cite{sigl1}.
 The origin of this primordial magnetic fields (PMF) can be searched  as electroweak and  QCD phase transitions, inflation, string theory, among others \cite{giovannini2}. Basically we can classify this seed in two groups depending  on generation
model (Inflation or post inflation scenarios).  If we consider an  inflation scenario for example,  we can get PMFs  on  scales  larger than the Hubble horizon with a  variety of spectral indices (supposing the power spectrum of PMF  has the form of a power law) \cite{sayantan}.  Whilst post inflationary  scenarios, causally PMFs are generated, thus the maximum coherence lenght for
the fields must be no less than  Hubble horizon  and also  the  spectral index is  equal or greater than two \cite{tina1}.  
If  PMFs really were present before to recombination era, these   could have some effect on  big bang nucleosynthesis (BBN), electroweak baryogenesis process and   would leaves imprints in the temperature and polarization anisotropies of the cosmic microwave background  (CMB)  \cite{sayantan}, \cite{Grasso}, \cite{giovannini-kunze,kunze2}.
This effect on CMB has been studied since the early attempts of Zeldovich  and nowadays it is a  subject of active investigation \cite{giovannini1}, \cite{lewis}, \cite{kunze1}, \cite{yamazaki1}, \cite{durrer1}.  
In cases where PMFs is supposing to be homogeneous, some authors have found these ones can produce effects on acustic peaks due to fast magnetosonic waves and Alfv\'en  waves  induce  correlations in temperature multipole moments \cite{tina1}, \cite{tina2}. Other alternative is to consider a stochastic PMF
where its power spectra is assumed to be  a power law. In this case the  Alfv\'en waves induced by  a stochastic magnetic field affect the  pattern of temperature and  B-polarization on CMB \cite{tina3}. Different works have addressed the study of PMFs in scenarios where these ones are  modeled via  stochastic fields because they are  more realistic and look like to the fields measured in clusters of galaxies  \cite{tina1}, \cite{Dreher}. Also, in  \cite{paoletti} they studied the impact of a stochastic PMF on scalar, vector and tensor modes on CMB anisotropies, finding  that the vector modes dominate over the scalar ones at high multipolar numbers and in \cite{riotto}, \cite{trivedi} is analysed the non-Gaussian signals on CMB   generated via stochastic PMFs. In this paper we focus our study in the case of PMFs generated in post inflationary stages and its influence in the CMB anisotropies. For this,  we  calculate the exact scalar, vector and tensor power spectrum for the energy density, Maxwell stress-energy tensor and Lorentz force of a  stochastic PMF with an upper cutoff at $k_D$  which corresponds to the damping scale and a lower cutoff $k_m$  which corresponds to the Hubble radius when the field was generated. 
Indeed, this $k_m$ gives the minimum wave numbers and it is dependent on PMF generation models, therefore this lower cutoff could give us information about the PMF
generation mechanisms, and thus its study  will be of great importance in this paper. We also calculate   the angular power spectrum of the CMB temperature anisotropy induced by a  magnetic perturbation.
This paper is organised as follows: Section 2 describes the  two-point correlation function for a statistically homogeneous and isotropic magnetic field, Section 3 explains the  cutoff in the definitions of the power spectrum, Section 3 presents  the integration technique and  Section 4 reports numerical solutions of the power spectrum of a PMF.   With the exact expression of the power spectrum,  the angular power spectrum of the CMB induced by PMFs  is computed in Section 5. Finally, a summary of the work and  conclusions are presented in Section 6.
 
\section{Magnetic correlation functions}
To deal with a PMF, the space-time under study is  permeated by a weak magnetic field, which is a stochastic field and can be treated as a perturbation on a flat-Friedman-Lemaitre-Robertson Walker (FLRW) background
\begin{equation}
 ds^{2}=a^{2}(\tau)\left(-d\tau^{2}+ \delta_{ij}dx^{i}dx^{j}\right), 
\end{equation}
with $a(\tau)$ the scale factor\footnote{ Hereafter the Greek indices run from 0 to 3, and the Latin ones run from 1 to 3, we will work with conformal time $\tau$,  $\tau_0$ is the current value of conformal time. 
}. 
 The  electromagnetic energy momentum tensor at first order in the perturbation theory is quadratic in the magnetic fields
\begin{eqnarray}
T_{00}^{(B)}(\textbf{x},\tau)&=&\rho_B(\textbf{x},\tau)=\frac{1}{8\pi}B^2(\textbf{x},\tau),\\
T_{ij}^{(B)}(\textbf{x},\tau)&=&\frac{1}{4\pi}\left[B_i(\textbf{x},\tau)B_j(\textbf{x},\tau)-\frac{1}{3}\delta_{ij}B^2(\textbf{x},\tau)\right],
\end{eqnarray}
also,  the anisotropic trace-free part  of the stress-energy tensor (spatial part of energy momentum tensor) of the magnetic field  takes the form
\begin{equation}
\Pi_{ij}(\textbf{x},\tau)=T_{ij(B)}(\textbf{x},\tau)+\frac{1}{3}\delta_{ij}\rho_B(\textbf{x},\tau).
\end{equation}
The PMF amplitude scales as $B^2(\textbf{x},t)=\frac{B^2(\textbf{x})}{a^{4}(\tau)}$  at larges scales  within the infinite conductivity limit which is a good approximation  before the decoupling epoch \cite{hortua}.   
\subsection{The statistics for a stochastic PMF}
Now, the PMF power spectrum  which is defined as the Fourier transform of the two points correlation  can be written as
\begin{equation}\label{PMFespectro}
\langle B_i^*(\textbf{k})B_j(\textbf{k}^\prime)\rangle =(2\pi)^3\delta^3(\textbf{k}-\textbf{k}^\prime)P_{ij}P_B(\left|\textbf{k}\right|),
\end{equation}
where $P_{ij}$ is a projector onto the transverse plane\footnote{Being $P_{ij}=\delta_{ij}-\frac{\textbf{k}_i\textbf{k}_j}{k^2}$,   where $P_{ij}P_{jk}=P_{ik}$ and $P_{ij}\textbf{k}_j=0$.},  $P_B(\left|\textbf{k}\right|)$ is the
PMF power spectrum and  where we use the Fourier transform conventions
\begin{eqnarray}
B_j(\textbf{x})&=&\int  \frac{d^3x}{(2\pi)^3}\exp{(-i\textbf{k}\cdot \textbf{x})}B_j(\textbf{k}), \nonumber \\
\delta(\textbf{k})&=&\int \frac{d^3x}{(2\pi)^3}\exp{(i\textbf{k}\cdot \textbf{x})}.
\end{eqnarray}
Since  $B$ is statistically homogeneous and isotropic,  the correlation  depends only on the distance $\left|\textbf{x}-\textbf{y}\right|$. We restrict our attention to the evolution of a causally generated  or post inflationary PMF  parametrized by the power law with index $n \geq 2$, with  an ultraviolet cutoff $k_{D}$ and the dependence of an infrared
cutoff $k_m$, thus  we consider that for  $k_{m} \leq k \leq k_{D}$  the power spectrum can be defined as
\begin{equation}\label{powerPMF}
P_B(k)=Ak^n,
\end{equation}
being $A$ the normalization constant which is given in \cite{tina1} as
 \begin{equation}
A=\frac{B^2_{\lambda}2\pi^2\lambda^{n+3}}{\Gamma(\frac{n+3}{2})},
\end{equation}
where $B_\lambda$ is the comoving PMF strength smoothing over a Gaussian sphere of comoving radius $\lambda$. The equations,  for  energy density of magnetic field and anisotropic trace-free part respectively  written in Fourier space are
\begin{eqnarray}\label{covdensity}
\rho_B(k,\tau)&=&\frac{1}{8\pi}\int \frac{d^3k^\prime}{(2\pi)^3}B_l(k)B^l(\left|\textbf{k}-\textbf{k}^\prime\right|),\\
\Pi_{ij}(k,\tau)&=&\frac{1}{4\pi}\int \frac{d^3k^\prime}{(2\pi)^3}\left[B_i(k^\prime)B_j(\left|\textbf{k}-\textbf{k}^\prime\right|) \right. \nonumber \label{convstress}\\
&-&  \frac{1}{3}\delta_{ij}B_l(k^\prime)B^l(\left|\textbf{k}-\textbf{k}^\prime\right|) \left. \right].
\end{eqnarray}
Following  \cite{kunze}, the anisotropic trace-free part can be splitted in a scalar, vector and tensor part
\begin{eqnarray}
\Pi^{(S)}(k,\tau)&=&\frac{3}{2}\left(\frac{\textbf{k}_i\textbf{k}_j}{k^2}-\frac{1}{3}\delta_{ij}\right)\Pi^{ij}(k,\tau),\\
\Pi_{i}^{(V)}(k,\tau)&=&P_{ij}\frac{\textbf{k}_l}{k}\Pi^{lj}(k,\tau),\\
\Pi_{ij}^{(T)}(k,\tau)&=& P_{iljm} P^{mn}P^{ls}\Pi_{ns}(k,\tau),
\end{eqnarray}
where  scale in the same way that energy density (infinite conductivity) of PMF like $\Pi_{ij}(k,\tau)=\frac{\Pi_{ij}(k,\tau_0)}{a^4(\tau)}$ and $P_{iljm}=\left(P_{il}P_{jm}-\frac{1}{2}P_{ij}P_{lm}\right)$. Furthermore, PMFs  affect motions of ionized baryons by the Lorentz force which is read as
\begin{equation}
\textbf{L}(k,\tau_0)=\frac{1}{4\pi}\left((\nabla \times \textbf{B}(\textbf{x}))\times \textbf{B}(\textbf{x})\right),
\end{equation}
which appears in the  Navier-stokes equation at first order when PMF is considered \cite{hortua}. Using the free divergence  of magnetic field property and the decomposition the Lorentz force into a scalar and  vector part, the relation between the  anisotropic  stress-energy tensor and Lorentz force is given by \cite{tina4}
\begin{equation}
\Pi^{(S)}(\textbf{x},\tau_0)=L^{(S)}(\textbf{x},\tau_0)+\frac{1}{3}\rho_B(\textbf{x},\tau_0).
\end{equation}
Now, we use the  two-point correlation function for  $\rho_B(k,\tau)$,  $\Pi(k,\tau)$,  $L_B(k,\tau)$ and the cross-correlation between them
\begin{equation}
\langle \rho_B(\textbf{k},\tau)\rho_B^{*}(\textbf{k}^\prime,\tau)\rangle=(2\pi)^3 \left|\rho_B(k,\tau)\right|^2\delta^3(\textbf{k}-\textbf{k}^\prime),
\end{equation}
\begin{equation}
\langle \Pi^{(S)}(\textbf{k},\tau)\Pi^{(S)*}(\textbf{k}^\prime,\tau)\rangle=(2\pi)^3 \left|\Pi^{(S)}(k,\tau)\right|^2\delta^3(\textbf{k}-\textbf{k}^\prime),
\end{equation}
\begin{equation}
\langle L^{(S)}(\textbf{k},\tau)L^{(S)*}(\textbf{k}^\prime,\tau)\rangle=(2\pi)^3 \left|L^{(S)}(k,\tau)\right|^2\delta^3(\textbf{k}-\textbf{k}^\prime),
\end{equation}
\begin{equation}
\langle \rho_B(\textbf{k},\tau)L^{(S)*}(\textbf{k}^\prime,\tau)\rangle=(2\pi)^3 \left|\rho_B(k,\tau)L^{(S)}(k,\tau)\right|\delta^3(\textbf{k}-\textbf{k}^\prime),
\end{equation}
\begin{equation}
\langle \rho_B(\textbf{k},\tau)\Pi^{(S)}(\textbf{k}^\prime,\tau)\rangle=(2\pi)^3 \left|\rho_B(k,\tau)\Pi^{(S)}(k,\tau)\right|\delta^3(\textbf{k}-\textbf{k}^\prime),
\end{equation}
for the scalar part.  For the vector and tensor part we have
\begin{equation}
\langle \Pi_{i}^{(V)}(\textbf{k},\tau)\Pi_{j}^{(V)*}(\textbf{k}^\prime,\tau)\rangle=(2\pi)^3 P_{ij}\left|\Pi^{(V)}(k,\tau)\right|^2\delta^3(\textbf{k}-\textbf{k}^\prime),
\end{equation}
\begin{equation}
\langle \Pi_{ij}^{(T)}(\textbf{k},\tau)\Pi_{ij}^{(T)*}(\textbf{k}^\prime,\tau)\rangle=4(2\pi)^3 \left|\Pi^{(T)}(k,\tau)\right|^2\delta^3(\textbf{k}-\textbf{k}^\prime),
\end{equation}
  respectively, here the power spectrum   depends only on $k = |\textbf{k}|$. Now, to calculate the power spectrum, we substitute the equations  (\ref{covdensity}) and (\ref{convstress}) in the above expressions, then we use the  Wick's theorem to evaluate the four-point correlator of the PMF and finally the   equation  (\ref{PMFespectro}) is used.  After a straightforward but somewhat lengthy calculation one obtains the power spectrum for  $\rho_B(k,\tau)$,  $\Pi(k,\tau)$,  $L_B(k,\tau)$ given by

\begin{eqnarray}
\left|\rho_B(k,\tau)\right|^2&=&\frac{1}{256\pi^5}\int d^3k^\prime(1+\mu^2)P_B(k^\prime)P_B(\left|\textbf{k}-\textbf{k}^\prime\right|),\\
\left|L^{(S)}(k,\tau)\right|^2&=&\frac{1}{256\pi^5}\int d^3k^\prime[4(\gamma^2\beta^2-\gamma\mu\beta)+1+\mu^2]\times \nonumber\\
&\times&P_B(k^\prime)P_B(\left|\textbf{k}-\textbf{k}^\prime\right|),\\
\left|\Pi^{(s)}(k,\tau)\right|^2&=&\frac{1}{576\pi^5}\int d^3k^\prime[4-3(\beta^2+\gamma^2)+\mu^2 \nonumber\\
&+&9\gamma^2\beta^2-6\mu\beta\gamma]P_B(k^\prime)P_B(\left|\textbf{k}-\textbf{k}^\prime\right|),
\end{eqnarray}
for scalar modes
\begin{eqnarray}
\left|\rho_B(k,\tau)L^{(S)}(k,\tau)\right|&=&\frac{1}{256\pi^5}\int d^3k^\prime[1-2(\gamma^2+\beta^2) \nonumber\\
&+& 2\gamma\mu\beta-\mu^2]P_B(k^\prime)P_B(\left|\textbf{k}-\textbf{k}^\prime\right|),\\
\left|\rho_B(k,\tau)\Pi^{(S)}(k,\tau)\right|&=&\frac{1}{128\pi^5}\int d^3k^\prime \left[\frac{2}{3}-(\gamma^2+\beta^2) \right. \nonumber\\
&+& \left. \mu\gamma\beta-\frac{1}{3}\mu^2\right]P_B(k^\prime)P_B(\left|\textbf{k}-\textbf{k}^\prime\right|), 
\end{eqnarray}
for the scalar  cross-correlation and
\begin{eqnarray}
\left|\Pi^{(V)}(k,\tau)\right|^2&=&\frac{1}{128\pi^5}\int d^3k^\prime[(1+\beta^2)(1-\gamma^2)\nonumber\\
&+&\mu\gamma\beta-\gamma^2\beta^2]P_B(k^\prime)P_B(\left|\textbf{k}-\textbf{k}^\prime\right|),\\
\left|\Pi^{(T)}(k,\tau)\right|^2&=&\frac{1}{512\pi^5}\int d^3k^\prime[1+2\gamma^2+\gamma^2\beta^2]\times\nonumber\\
&\times&P_B(k^\prime)P_B(\left|\textbf{k}-\textbf{k}^\prime\right|),
\end{eqnarray}
for the vector and tensor part. The angular functions are defined as
\begin{equation}
\beta=\frac{\textbf{k}\cdot(\textbf{k}-\textbf{k}^\prime)}{k\left|\textbf{k}-\textbf{k}^\prime\right|}, \quad \mu=\frac{\textbf{k}^\prime\cdot(\textbf{k}-\textbf{k}^\prime)}{k^\prime\left|\textbf{k}-\textbf{k}^\prime\right|}, \quad  \gamma=\frac{\textbf{k}\cdot\textbf{k}^\prime}{kk^\prime}.
\end{equation}
Our results are in agreement with those found by \cite{tina4}, \cite{paoletti1}, \cite{lewis}.

\section{The cutoff dependence with the scale}
In this part we solve the last expressions for getting the power spectrum of a causal PMF generated before recombination epoch.
By considering a stochastic PMF in the cosmological scenario,  an upper cutoff $k_D$ corresponds to the damping scale should be taking in account, in sense that magnetic field energy is dissipated into heat through the damping of magnetohydrodynamics waves. The damping ocurrs due to the diffusion of  neutrinos prior to neutrino decoupling (T $\sim 1$MeV) and  the photons before recombination (T $\sim 0.25$ eV). Particularly, we try with three types of propagating MHD modes, the fast and slow magnetosonic waves  and the Alfv\'en waves \cite{tina1}, \cite{Jedamzik1}.
However, we concentrate in the latter because these ones are the most effective in damping when radiation is free-streaming (recombination), that is, when 
$k \leq V_AL_{Silk}$, where $V_A$ is the Alfv\'en speed and $L_{Silk}$, the Silk damping scale at recombination, \cite{barrow}, \cite{riotto}.   
The upper cutoff of PMF was found by \cite{barrow}, \cite{tina1} which  is dependent of strength  of magnetic energy and the spectral index as follows
\begin{eqnarray}
\frac{k_D}{h^{\frac{1}{n+5}}Mpc^{-1}}&\approx&\left(1.7\times 10^2\right)^{\frac{2}{n+5}}\left(\frac{B_\lambda}{1nG}\right)^{\frac{-2}{n+5}}\nonumber\\
&\times&\left(\frac{k_\lambda}{1Mpc^{-1}}\right)^{\frac{n+3}{n+5}},
\end{eqnarray}
for vector modes where $h=0.679\pm 0.100$.  For tensor modes the cutoff takes the following form 
\begin{eqnarray}
\frac{k_D}{h^{\frac{6}{n+5}}Mpc^{-1}}&\approx&\left(8.3\times 10^3\right)^{\frac{2}{n+5}}\left(\frac{B_\lambda}{1nG}\right)^{\frac{-2}{n+5}}\nonumber\\
&\times&\left(\frac{k_\lambda}{1Mpc^{-1}}\right)^{\frac{n+3}{n+5}}.
\end{eqnarray}
Therefore,  the damping scale changes with time and the power spectrum for a PMF  must  have a  time dependence due to the cosmic epoch where it is present  besides the decay by the expansion of the universe.
With the latter equations, we can expect a high contribution of tensor modes on CMB for large scales respect to the vector ones.   
Now, the  power  spectrum of magnetic field  we want to study takes into account an infrared cutoff $k_m$  for low values
of $k$ and  which  depend on the generation model of the PMF. 
This minimal scale has been studied by \cite{yamazaki3},  \cite{yamazaki2}, \cite{yamazaki4}, \cite{olinto}  showing the effects of PMFs on abundances of primordial light elements using BBN, the distortions on CMB due to a  background PMF and the relevance of PMF in formation of structure in the universe respectively. Therefore,  the scale $k[\mbox{Mpc}^{-1}]$ moves from $k_m$ to $k_D$ and where we parametrize this infrared cutoff as $k_m= \alpha k_D$ where $0<\alpha<1$. 
This lower cutoff is strongly dependent on the PMF generation model. Therefore, studying its effects  on  the CMB signal we could get information about the PMF generation mechanism. 
\subsection{Integration method}
We choose our coordinate system in such a way that $\textbf{k}$ is along the $\textbf{z}$ axis, thus $\gamma$ is the cosine of angle between $\textbf{k}^\prime$ and the $\textbf{z}$ axis ($\gamma=cos \theta$). The integration measure can be written, in spherical coordinates as $d^3k^\prime=k^{\prime \, 2}dk^\prime d\gamma d\phi$. The angular part related with $\phi$ is just equal to $2 \pi$. But, there is a constraint on the angle to be integrated over $\gamma$, depending on the magnitude of $\textbf{k}^\prime$. For making the integration  two conditions need to be fulfill
\begin{equation}\label{cond}
k_m<\left|\textbf{k}-\textbf{k}^\prime\right|<k_D, \quad k_m<\left|\textbf{k}^\prime\right|<k_D.
\end{equation}
Under these conditions, the power spectra is  non zero only for $0 < k < 2k_D$, result also found in \cite{paoletti}. The integration domain for calculating the power spectrum is found in  appendix \ref{appendix}. 
\section{Post inflationary magnetic field power spectra}
\begin{figure}[h!]
  \centering
    \includegraphics[width=.48\textwidth]{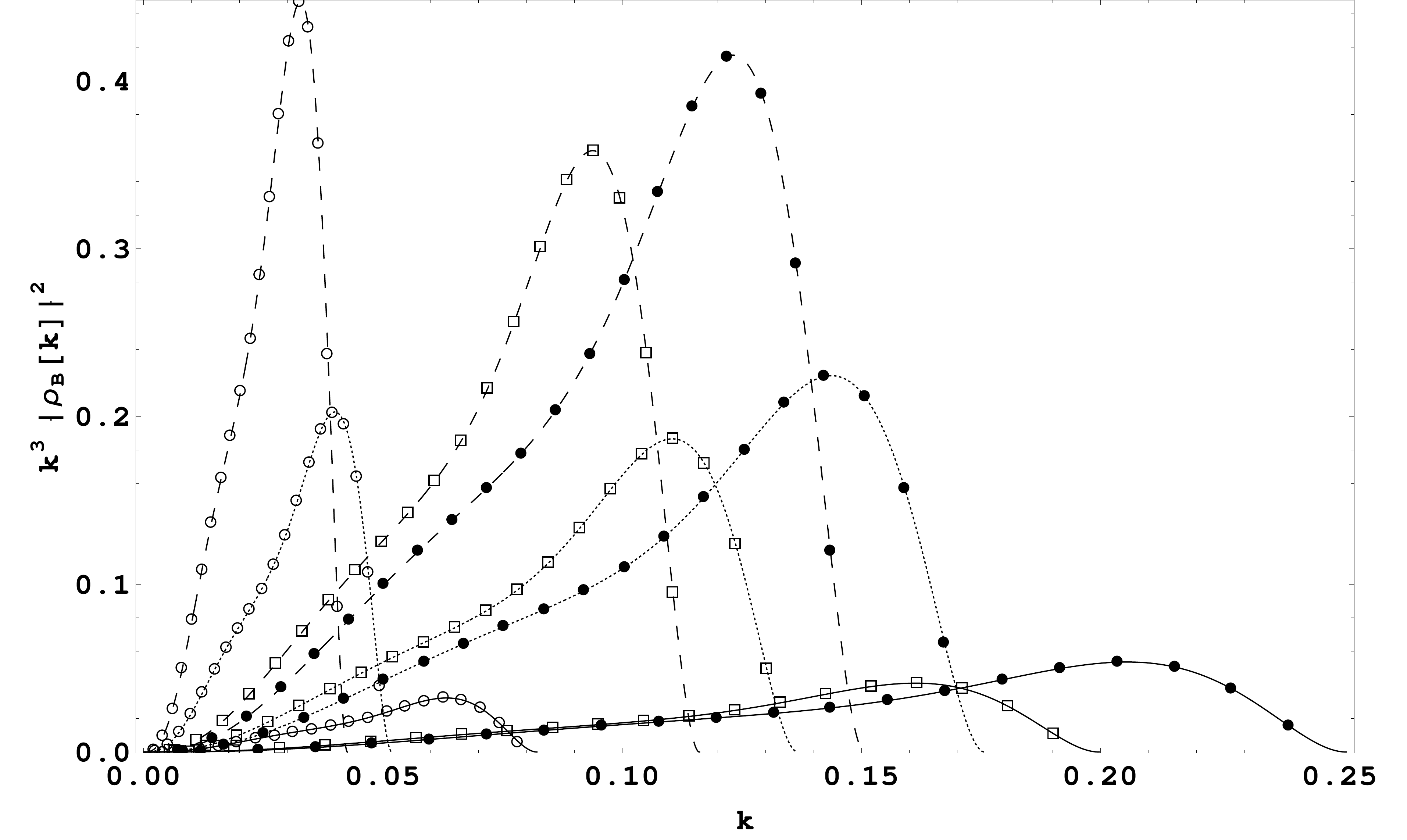}
  \caption{Plot of magnetic energy density of PMF power spectrum $k^3\left|\rho_B(k,\tau)\right|^2$ versus $k(\mbox{Mpc}^{-1})$ for different strenght of the PMF  (solid lines for $B_\lambda=1$nG, medium dashed lines for $B_\lambda=5$nG, and large dashed lines for $B_\lambda=10$nG), and for different spectral indices ($n=2$ for lines with open circles, $n=7/2$ for lines with open squares,  and $n=4$ for lines with filled circles).}
  \label{densidad1}
\end{figure}

\begin{figure}[h!]
  \centering
    \includegraphics[width=.48\textwidth]{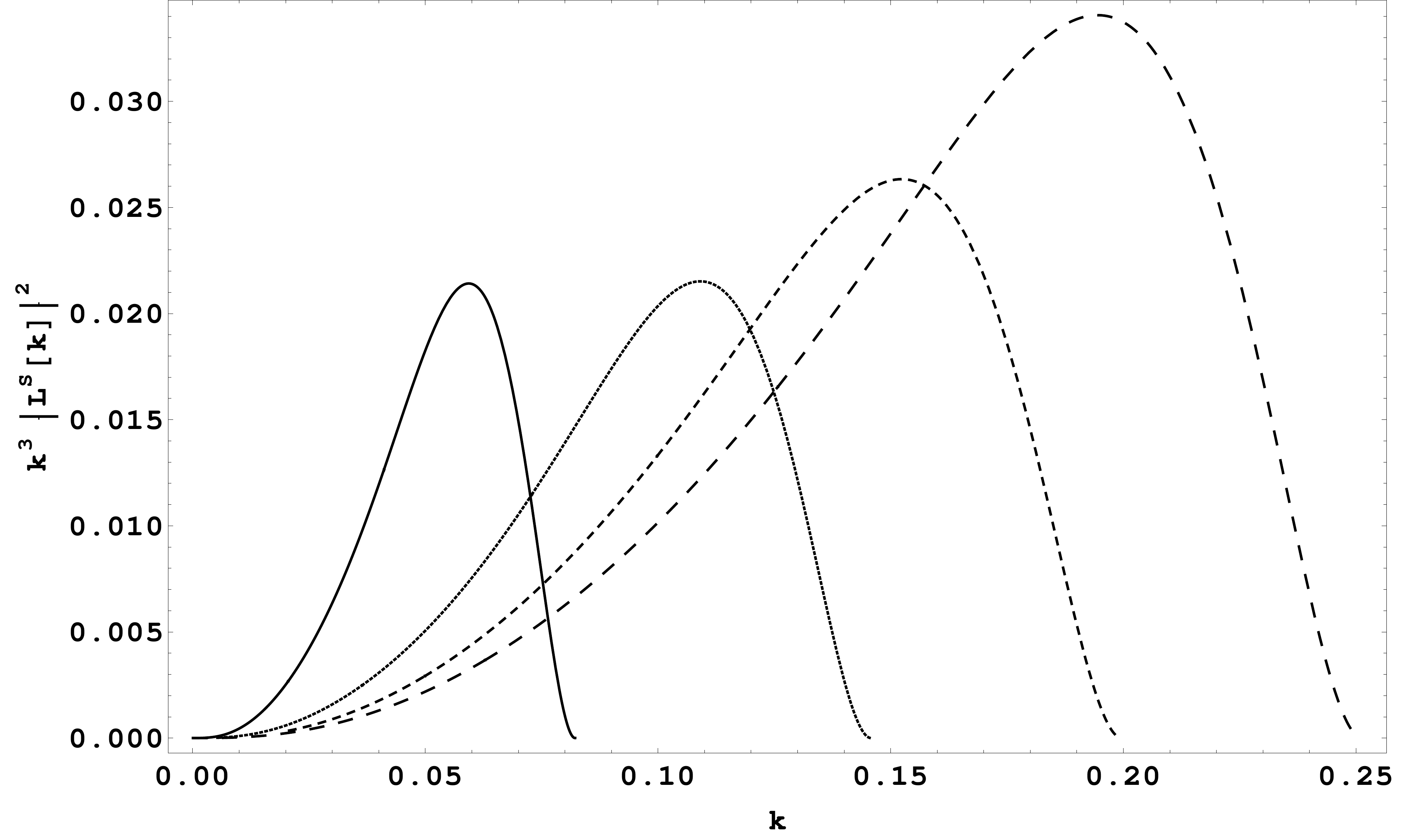}
  \caption{Plot of Lorentz force spectra $k^3\left|L^{(S)}(k,\tau)\right|^2$ versus $k(\mbox{Mpc}^{-1})$ for  different spectral indices ($n=2,3,7/2,4$ from solid to the large dashed lines).}
  \label{lorentz1}
\end{figure}

\begin{figure}[h!]
  \centering
    \includegraphics[width=.48\textwidth]{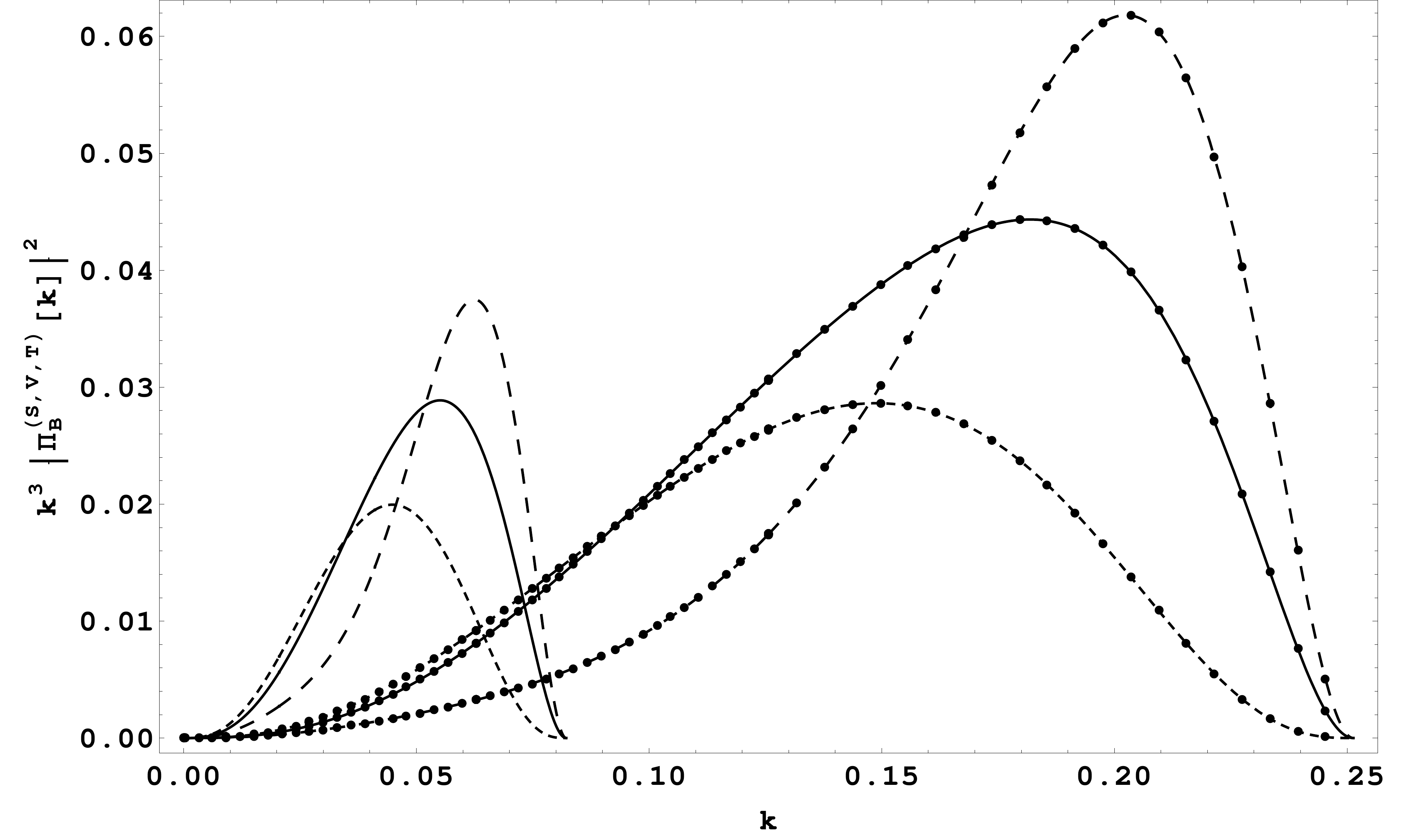}
  \caption{Plot of scalar (solid lines), vector (short dashed lines), and tensor (large dashed lines) parts of the anisotropic trace-free part power spectrum $k^3\left|\Pi(k,\tau)\right|^2$ versus $k(\mbox{Mpc}^{-1})$ for  different spectral indices ($n=2$ for lines  and $n=4$ for lines with filled circles). }
  \label{maxwell}
\end{figure}

Primordial magnetic fields generated after inflation are expected to have a very small amplitude ($10^{-20}$G) at the scale of 1Mpc, but even if this field is very small it is nonzero and it can leave a detectable imprint  on CMB pattern \cite{tinaver1}, \cite{tinaver2}.
The figure \ref{densidad1}  shows the magnetic energy density convolution and its dependence with both the spectral index  and the  amplitude  at a scale of $\lambda = 1$Mpc (we plot the power spectra times $k^3$ for comparing with \cite{paoletti1}).   We note that  amplitude of the spectra is proportional not only with the strenght of PMF as well to spectral index. 
In the figure \ref{lorentz1} the Lorentz force spectra is shown for different values of spectral index  keeping an amplitud of 1nG at a scale of 1Mpc. The scalar, vector and tensor anisotropic modes are shown in figure \ref{maxwell}. 
In this plot we can see that the largest contribution comes from tensor modes followed by scalar and  vector modes respectively.
\begin{figure}[h!]
  \centering
    \includegraphics[width=.48\textwidth]{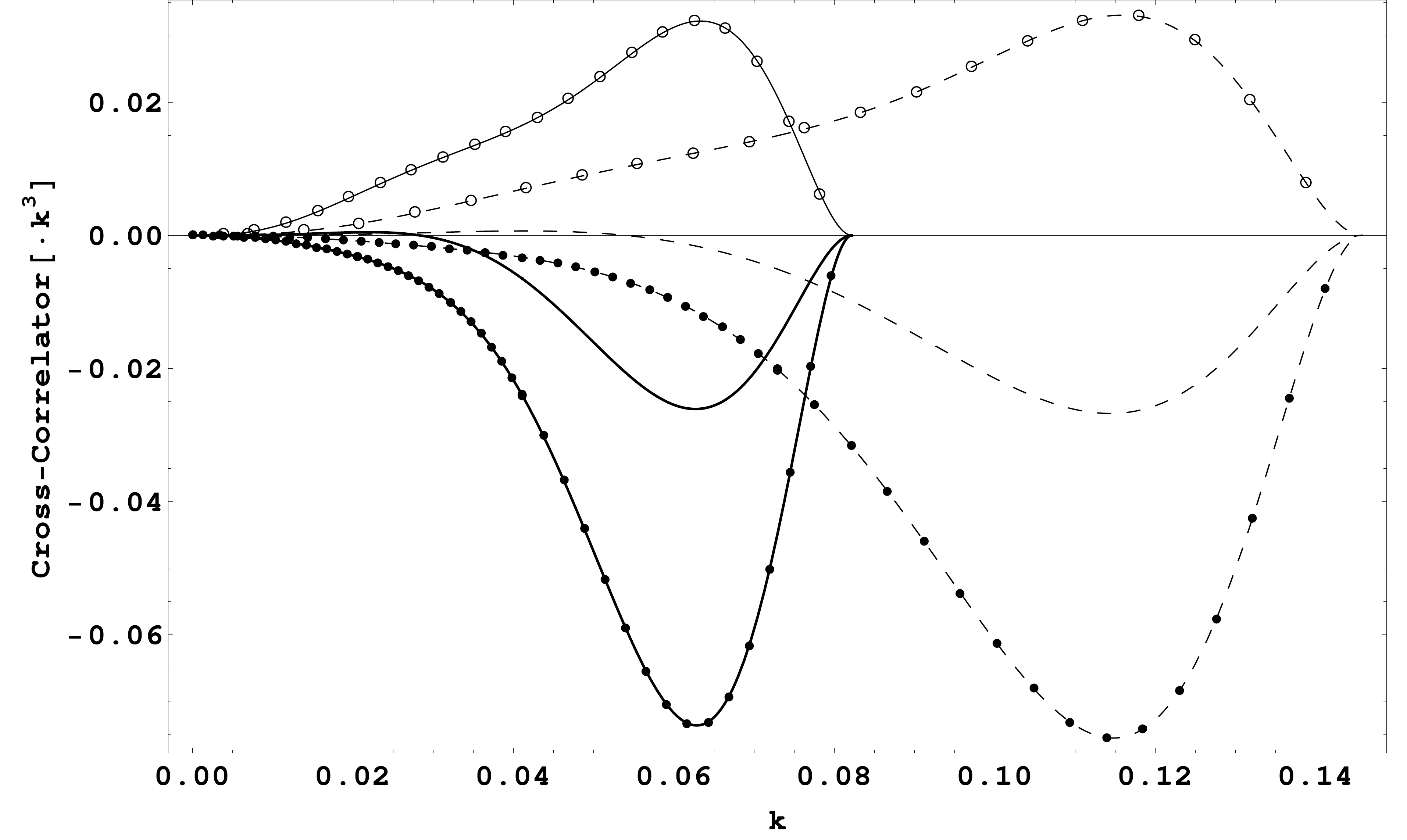}
  \caption{Plot of cross-correlation of $k^3\left|\rho_B(k,\tau)\Pi^{(S)}(k,\tau)\right|$ (just lines),   $k^3\left|\rho_B(k,\tau)L^{(S)}(k,\tau)\right|$ (lines with filled circles), and $k^3\left|\rho_B(k,\tau)\right|^2$ (lines with open circles),  for different spectral indices ($n=2$ for solid lines and  $n=3$ for dashed lines).}
  \label{correlators}
\end{figure}
The figure \ref{correlators}  shows  the cross correlation between the energy density with  Lorentz force and anisotropic trace-free part.  Notice that  the cross correlation between energy density and  Lorentz force is negative in all range of scales
 whilst the cross correlation between energy density and anisotropic trace-free part  starts to be negative for values of $k \geq 0.05 \mbox{Mpc}^{-1}$ (with $n=3$) and $k \geq 0.03\mbox{Mpc}^{-1}$ (with $n=2$).
The effect of the  smoothing scale over power spectrum is shown in the figures  \ref{varlambda} and \ref{varlambdat}, where we set the strengh of the field to 1nG.
The figure \ref{comparison}  makes  a comparison of our results of vector and tensor anisotropic trace-free parts with the found by \cite{paoletti} (see figure 1 and equation (A2) in this paper) and by \cite{tina1} (see equations (2.18), (2.22) in this paper) with  values of $B_{\lambda}=1$nG, $n=2$, at $\lambda=1$Mpc. Our results are in complete agreement  with the first authors and are in concordance with the second author just for $k \leq 0.015\mbox{Mpc}^{-1}$ for vector modes and  $k \leq 0.005\mbox{Mpc}^{-1}$ for tensor modes  and with a small difference in the  amplitude of the field. The scale $k$ for this plot runs from $0$ to $k_D\sim 0.04\mbox{Mpc}^{-1}$ due to  the approximation  found by \cite{tina1} is valid only for this range. 
As a general result we should notice that  there is a strong dependence of power spectrum and the upper cutoff with variables such as field amplitude and spectral index. Indeed, we observe how the  increase in the strenght of the PMF moves the peak of the spectrum and the value of $k_m$ to large scales (lower $k$), different from what happens with the spectral index which shifts the peak of the spectra and  $k_m$ to high values of the $k$-scale. This behavior is similar with the smoothing scale where for high values of $\lambda$ the peak moves to lower values of $k$.  In this way, some authors refer this upper  cutoff  to be a free parameter which is dependent on the PMF generation model being very important to constraint magnetogenesis models and could be contrasted with damping PMFs scenarios \cite{yamazaki1}.   
\begin{figure}[h!]
  \centering
    \includegraphics[width=.48\textwidth]{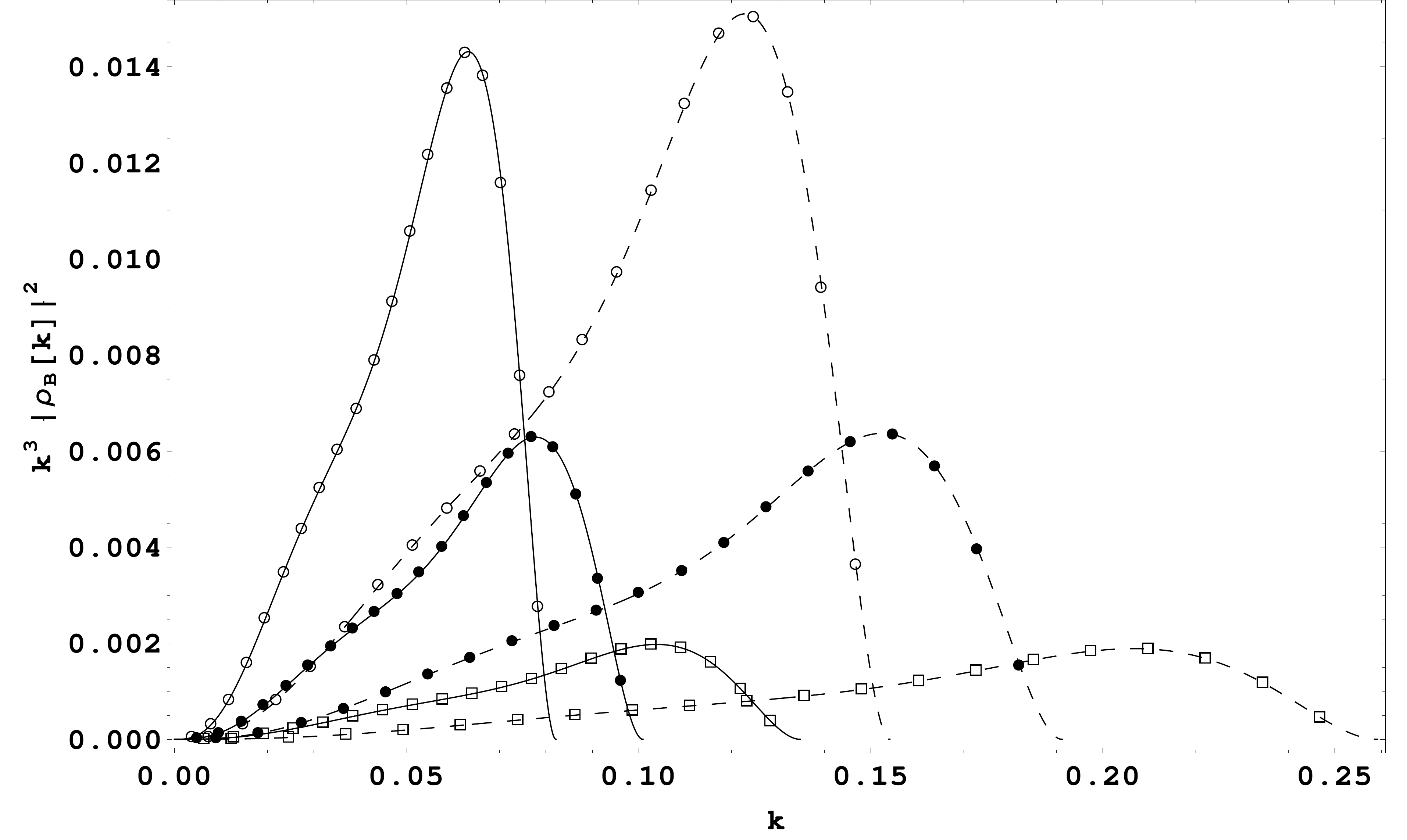}
  \caption{Plot of magnetic energy density of PMF power spectrum $k^3\left|\rho_B(k,\tau)\right|^2$ versus $k(\mbox{Mpc}^{-1})$ for different values of  smoothing scale ( $\lambda=1$Mpc for lines with open circles,  $\lambda=0.75$Mpc corres\-ponds to lines with filled circles, and  $\lambda=0.5$Mpc for lines with open squares),  and for  different spectral indices ($n=2$ refers to  solid lines and  $n=3$ for dashed lines).}
  \label{varlambda}
\end{figure}   
\begin{figure}[h!]
  \centering
    \includegraphics[width=.48\textwidth]{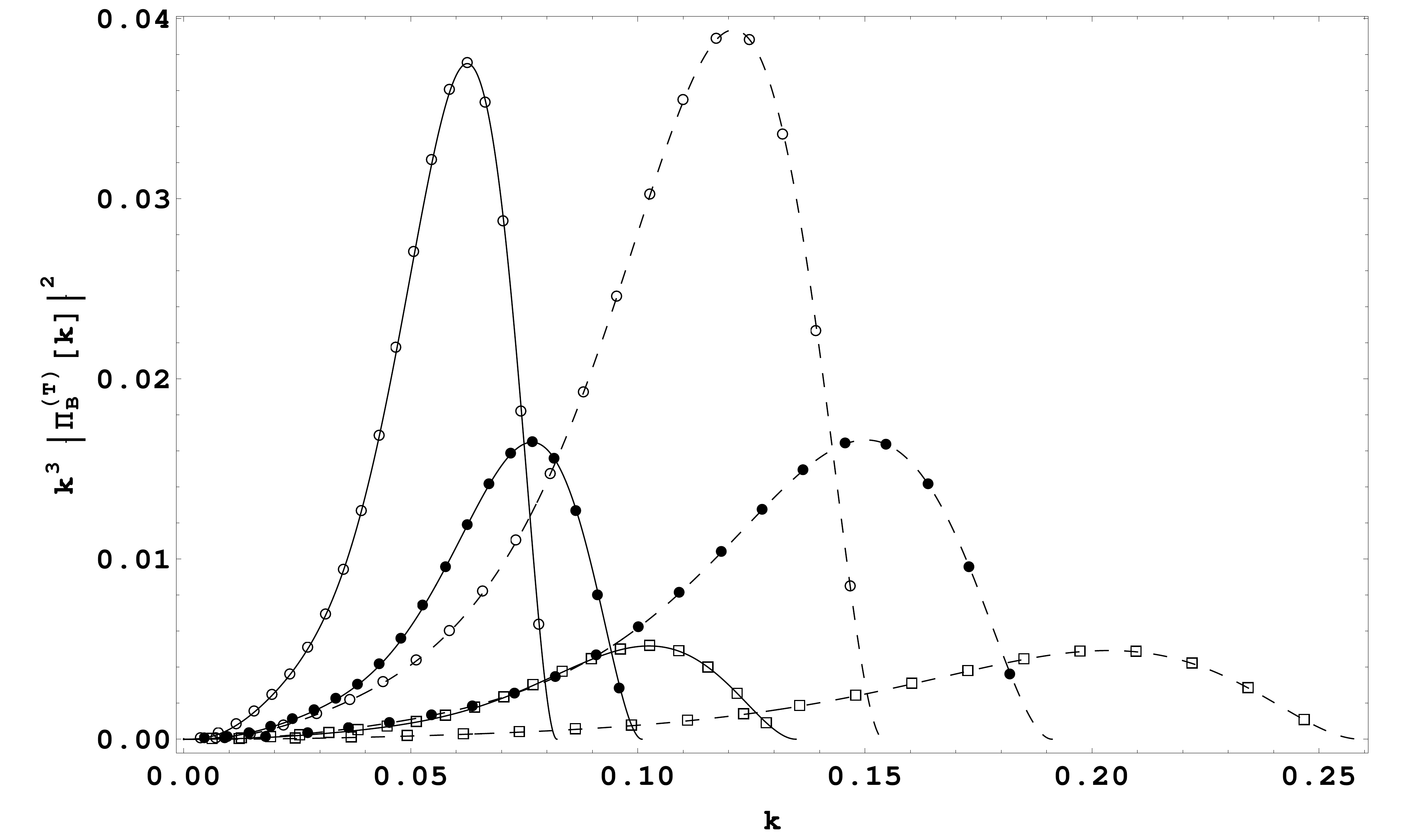}
  \caption{Plot of anisotropic trace-free tensor part power spectrum $k^3\left|\Pi^{(T)}(k,\tau)\right|^2$ versus $k(\mbox{Mpc}^{-1})$ for different values of  smoothing scale ( $\lambda=1$Mpc for lines with open circles,  $\lambda=0.75$Mpc corres\-ponds to lines with filled circles, and  $\lambda=0.5$Mpc for lines with open squares),  and for  different spectral indices ($n=2$ refers to  solid lines and  $n=3$ for dashed lines). }
  \label{varlambdat}
\end{figure}
\begin{figure}[h!]
  \centering
    \includegraphics[width=.48\textwidth]{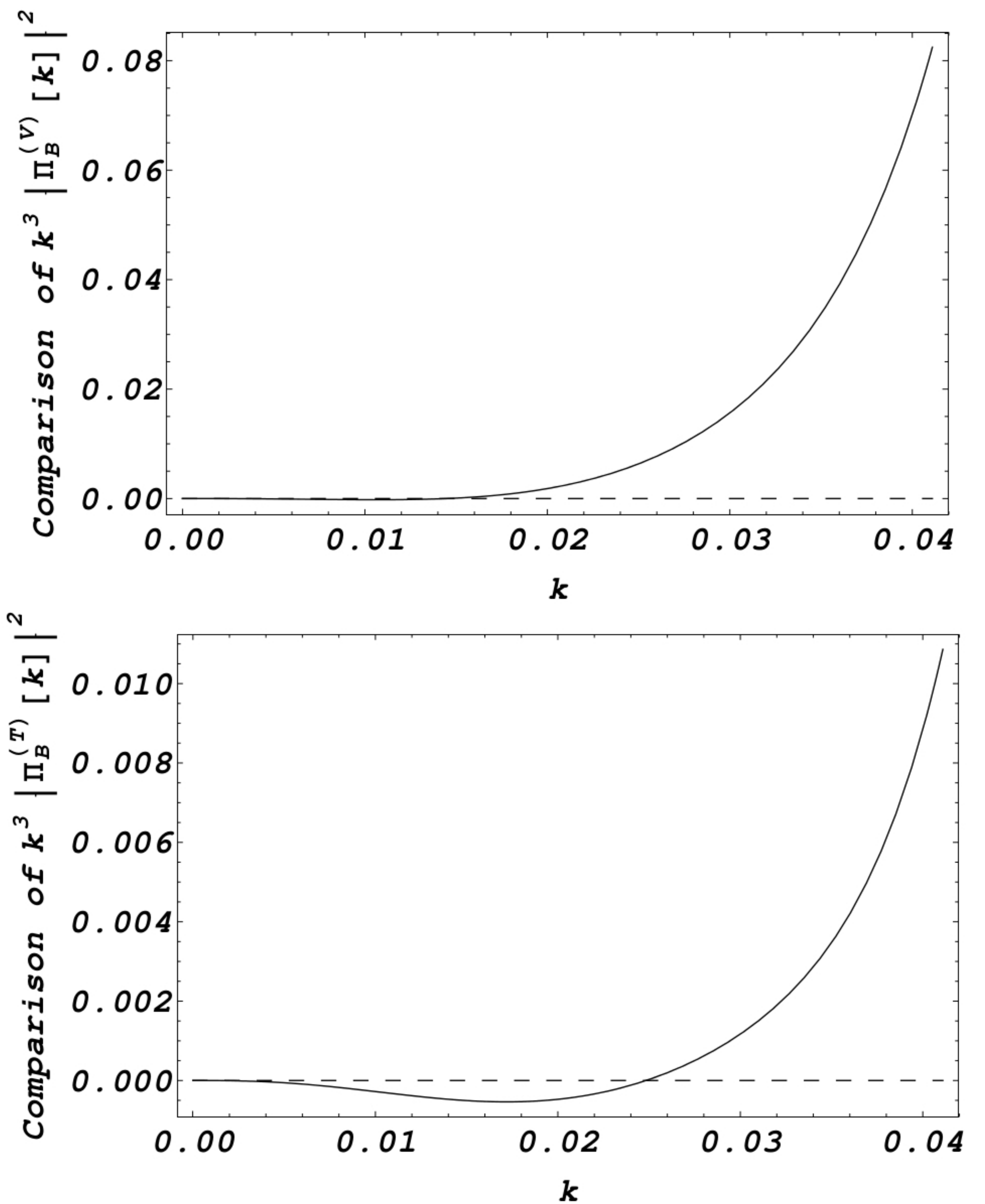}
  \caption{Plot of comparison between our results of the anisotropic trace-free part power spectrum $k^3\left|\Pi^{(V,T)}(k,\tau)\right|^2$ versus  \cite{paoletti} (dashed line) and \cite{tina1} (solid line) for values of $n=2$, $B_\lambda=1$nG and $\lambda=1$Mpc. }
  \label{comparison}
\end{figure}
\section{Magnetic  contribution to CMB anisotropies}
\begin{figure}[h!]
  \centering
    \includegraphics[width=.48\textwidth]{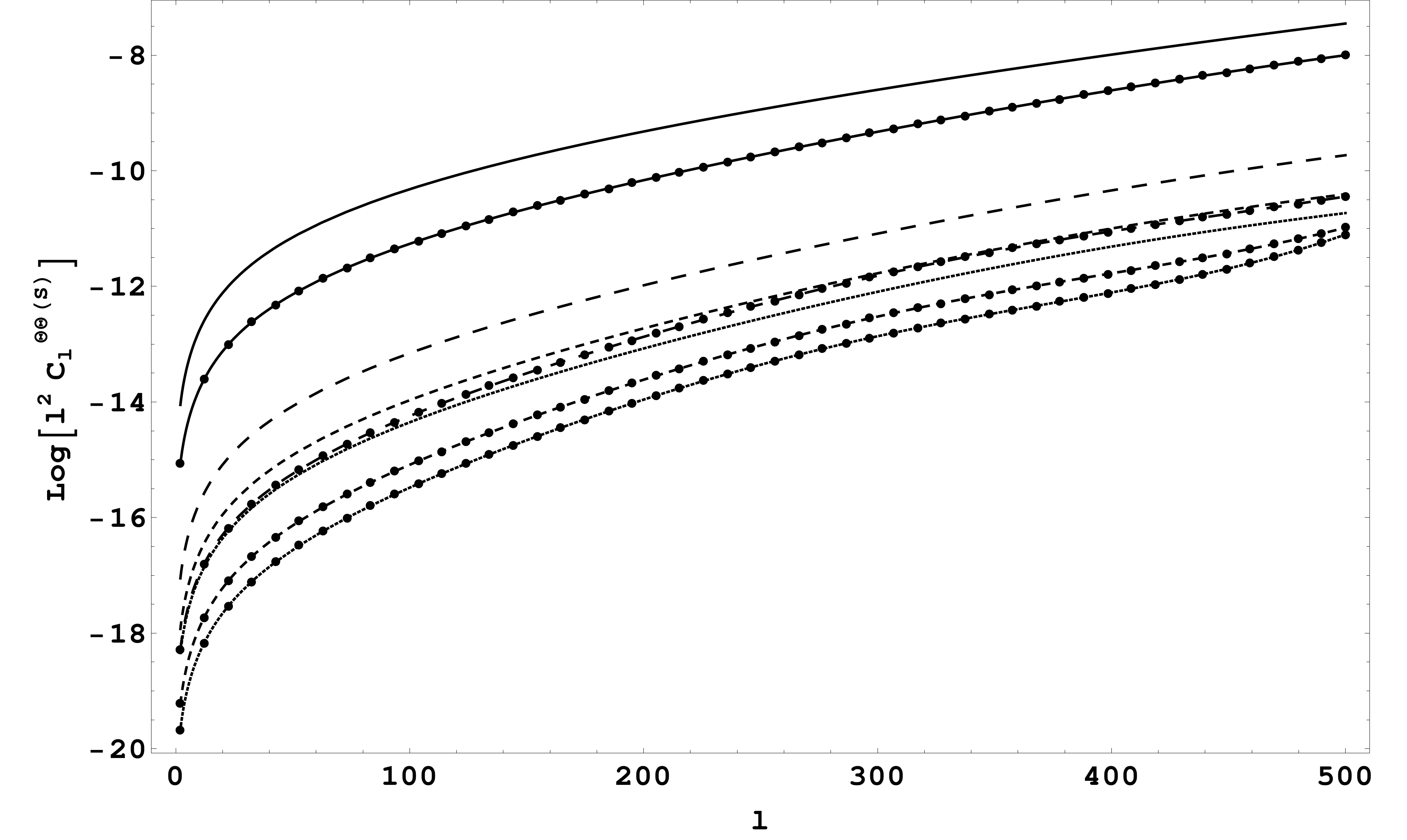}
  \caption{Plot of the CMB temperature anisotropy
angular power spectrum  induced by scalar
magnetic perturbations, where the  lines with filled circles  are for $n=2$ and the other ones for $n=5/2$. Here, the solid lines refer to $B_\lambda=10$nG, large dashed lines for
$B_\lambda=8$nG, small dashed lines refer to $B_\lambda=5$nG, and dotted lines for $B_\lambda=1$nG.}
  \label{cL1}
\end{figure}

Using the total angular momentum formalism introduced by \cite{hu}, the angular power spectrum of the CMB temperature anisotropy  is given as
\begin{equation}\label{momenttemp}
(2l+1)^2C_l^{\Theta \, \Theta}=\frac{2}{\pi}\int \frac{dk}{k}\sum_{m=-2}^2k^3 \Theta_l^{(m)\, *}(\tau_0,k)\Theta_l^{(m)}(\tau_0,k),
\end{equation}
where $ m = 0, \pm 1, \pm 2$ are the scalar, vector and tensor perturbations modes and  $\Theta_l^{(m)}(\tau_0,k)$ are the temperature fluctuation $\frac{\delta T}{T}$ multipolar moments.
In large scales,  one can neglect the contribution on CMB temperature anisotropies by ISW effect in presence of a PMF \cite{tina1}. Therefore, considering just the fluctuation via PMF perturbation,  the  temperature anisotropy multipole moment for $m=0$ becomes \cite{tina1}
\begin{equation}
\frac{\Theta_l^{(S)}(\tau_0,k)}{2l+1}\approx \frac{-8\pi G}{3k^2 a_{dec}^2}\rho_B(\tau_0,k)j_l(k\tau_0),
\end{equation}
where $a_{dec}$ is the value of scalar factor at decoupling, $G$ is the Gravitational constant and  $j_l$ is the spherical Bessel function.  Substituting the last expression  in equation (\ref{momenttemp}), the CMB temperature anisotropy
angular power spectrum is given by
\begin{equation}\label{eqcls}
l^2C_l^{\Theta \, \Theta\,(S)}=\frac{2}{\pi}\left(\frac{8\pi G}{3a_{dec}^2}\right)^2\int_0^\infty\frac{\left|\rho_B(\tau_0,k)\right|^2}{k^2}j_l^2(k\tau_0)l^2dk,
\end{equation}
where for our case,  we should integrate only up to  $2k_D$ since it is the range where energy density power spectrum is not zero.  The result of the angular power spectrum  induced by scalar magnetic perturbations given by equation (\ref{eqcls}) is shown in the figure \ref{cL1}.
\begin{figure}[h!]
  \centering
    \includegraphics[width=.48\textwidth]{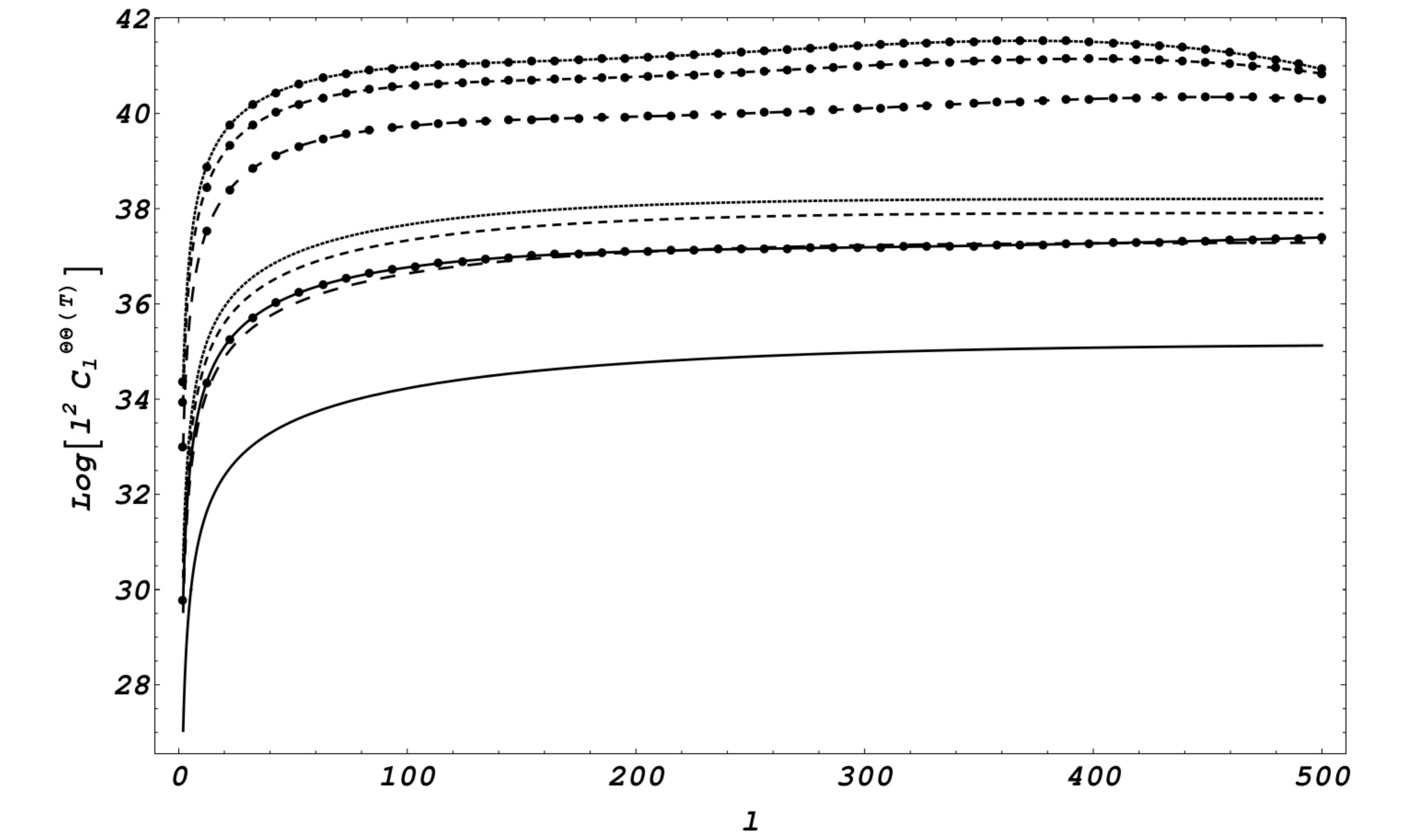}
  \caption{Plot of the CMB temperature anisotropy
angular power spectrum  induced by tensor
magnetic perturbations, where the  lines with filled circles are for $n=2$ and the other ones for $n=4$. Here, the solid lines refer to $B_\lambda=1$nG, large dashed lines for $B_\lambda=5$nG, small dashed lines refer to $B_\lambda=8$nG, and dotted lines for $B_\lambda=10$nG.}
  \label{clsT}
\end{figure}
 Here, we plot the $\log{l^2 C_l^{\Theta\Theta}}$ in order to compare our results with those found by \cite{tina1}. We calculate the angular power spectrum of  CMB in units of $\frac{2}{\pi}\left(\frac{8\pi G}{3a_{dec}^2}\right)^2$. One of the important features of the CMB power spectrum  (scalar mode) with a PMF is that distortion is proportional to strength of PMF and decreases with the spectral index and we must expect its  greatest contribution at low multipoles.  

In the case where $m \pm 2$ (tensor modes), the temperature anisotropy multipole moment  is given by Eq. (5.22) of \cite{tina1}
\begin{eqnarray}
\frac{\theta_l^{(T)}}{2l+1}&\simeq& -2\pi \sqrt{\frac{8(l+2)!}{3(l-2)!}}\left(G \tau_0^2z_{eq} \ln\left(\frac{z_{in}}{z_{eq}} \right )\right) \nonumber\\
&\times& \Pi^{(T)}(k,\tau_0)\int_0^{x_0}\frac{j_2(x)}{x}\frac{j_l(x_0-x)}{(x_0-x)^2}dx, 
\end{eqnarray}
where $z_{in}$ and $z_{eq}$ are the redshift  when PMF was created and during  equal matter-radiation era respectively  and $x_0=k\tau_0$. For the integral found in the last expression, we use the approximation made by \cite{tina2}
\begin{equation}
\int_0^{x_0}\frac{j_2(x)}{x}\frac{j_l(x_0-x)}{(x_0-x)^2}dx\simeq \frac{7 \pi}{25}\frac{\sqrt{l}}{x_0^3}J_{l+3}(x_0),
\end{equation}
where $j_l(z)=\sqrt{\frac{\pi}{2z}}J_{l+\frac{1}{2}}(z)$, being $J_{\nu}(z)$ the  Bessel functions of the first kind.  With this approximation the  tensor CMB temperature anisotropy angular power spectrum induced by a PMF  is given by
\begin{eqnarray}
l^2C_l^{\Theta \Theta (T)}&=&\left(G z_{eq} \ln\left(\frac{z_{in}}{z_{eq}} \right )\right)^2 \frac{l^4(l-1)(l+1)(l+2)}{(2l+1)^2 \tau_0^2}\nonumber \\
&\times& 1.25\pi^3\int\frac{dk}{k^4}J_{l+3}^2(k\tau_0)\left|\Pi^{(T)}(k,\tau_0) \right|^2.
\end{eqnarray}
The plot of CMB power spectra for  tensor  perturbations from a power law stochastic PMF  with spectral index $n=2$ ( lines with filled circles) and $n=4$ (without circles) for different amplitudes of the magnetic field is shown in figure \ref{clsT}. Here we can see the same dependence of spectral index and amplitud of PMF as the scalar case.
Here the spectra is in units of $\left(G z_{eq} \ln\left(\frac{z_{in}}{z_{eq}} \right )\right)^2\frac{1.25\pi^3}{\tau_0^2}$. 
 \section{Dependence of the spectrum with the infrared cutoff}
Studying the  effect of this lower cutoff of CMB spectra  we can constrain PMF generation models.
\begin{figure}[h!]
  \centering
    \includegraphics[width=.48\textwidth]{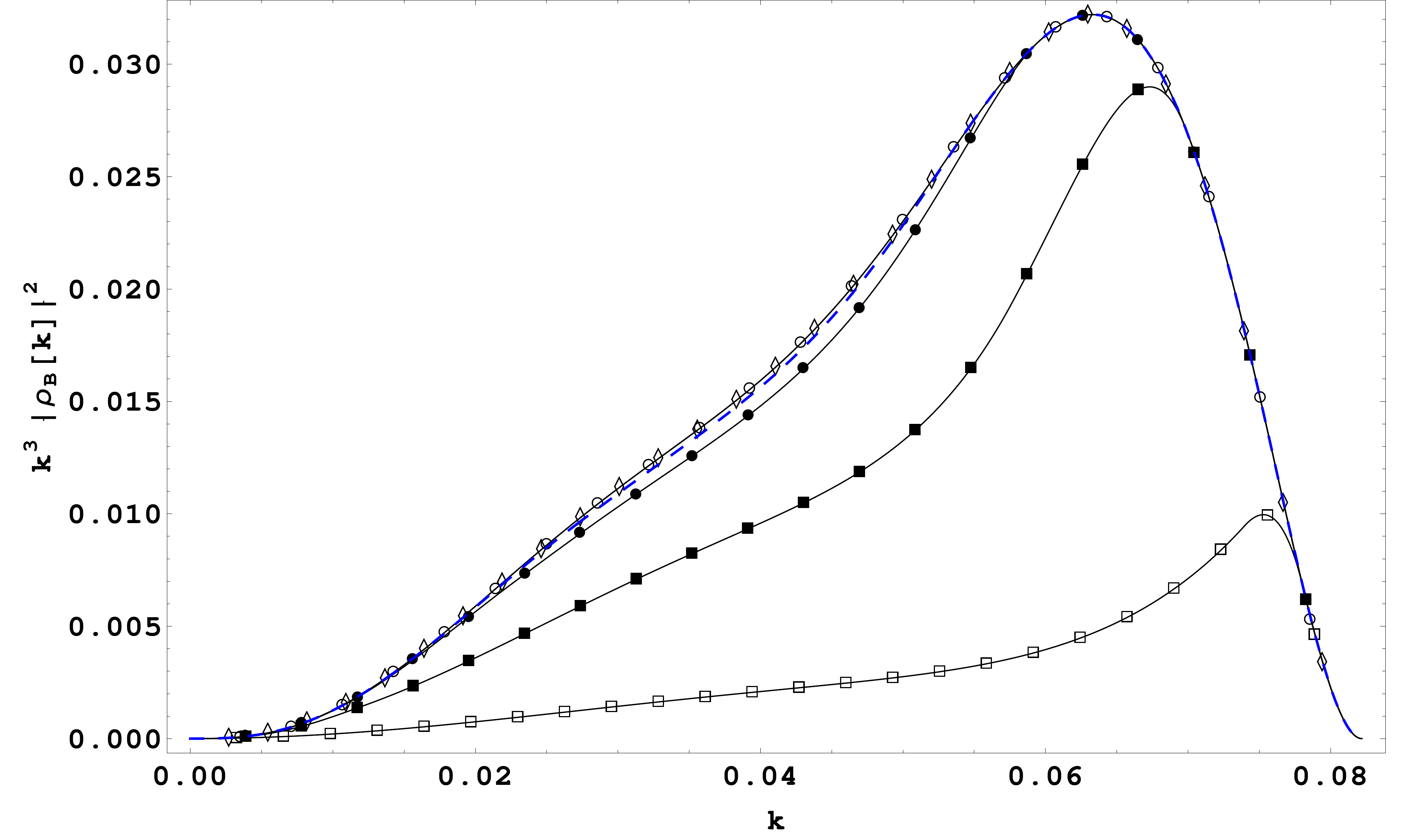}
  \caption{Plot of magnetic energy density of PMF power spectrum $k^3\left|\rho_B(k,\tau)\right|^2$ versus $k(\mbox{Mpc}^{-1})$ with $n_B=2$ and for different values of infrared cutoff, lines with open squares refer to  $k_m=0.9k_D$. Lines with filled squares correspond to  $k_m=0.7k_D$, lines with filled circles for $k_m=0.5k_D$; dashed line refer to $k_m=0.3k_D$, and finally, lines with open circles and diamonds correspond to $k_m=0.1k_D$ and  $k_m=0.001k_D$ respectively. }
  \label{variation}
\end{figure}
 For this, we plot in figure \ref{variation} the power spectrum of the energy density of PMF for different values of $k_m$. Here we can see the strong dependence of the power spectrum with this scale, basically the power spectrum does not change when  $0.2k_D>k_m>0$ with respect to the results of $k_m=0$, but in the cases where $k_m>0.2k_D$ (threshold described by dashed line) there is a significant variation with a null lower cutoff. Futhermore, for $k$ close to $2k_D$ the spectrum decays with the same slope, independent from lower cutoff, in this case for $n_B$, the slope of the energy density of PMF goes as $\sim k^{-3.2}$.
\begin{figure}[h!]
  \centering
    \includegraphics[width=.48\textwidth]{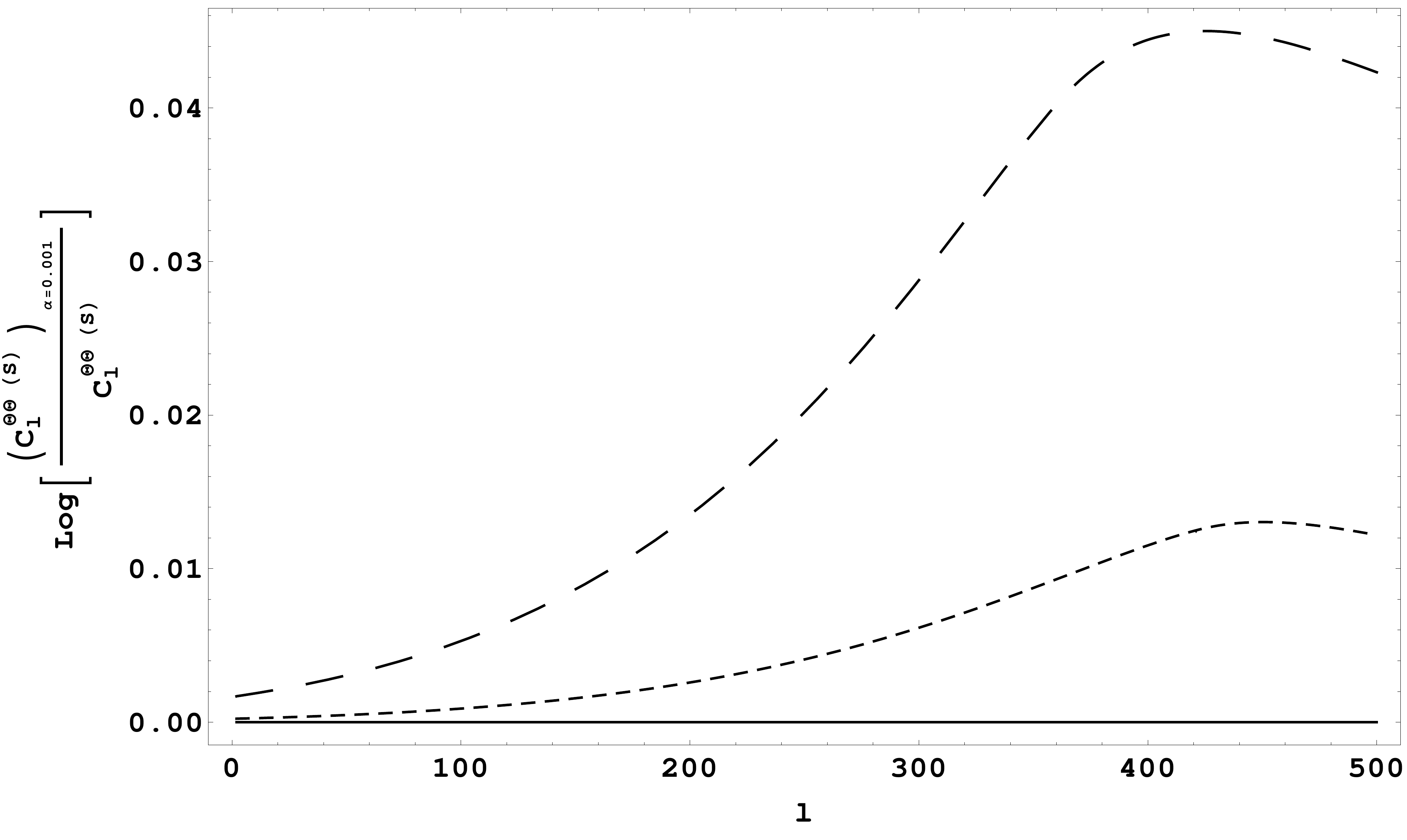}
  \caption{Comparison between the  CMB temperature anisotropy
angular power spectrum  induced by scalar PMF at $k_m=0.001k_D$ lower cutoff,  respect to the other ones with different values of infrared cutoff. Here, the solid horizontal line is for $k_m=0.1k_D$; small and large dashed  lines refer to $k_m=0.3k_D$ and  $k_m=0.4k_D$ respectively. }
  \label{clsT1}
\end{figure}
\begin{figure}[h!]
  \centering
    \includegraphics[width=.48\textwidth]{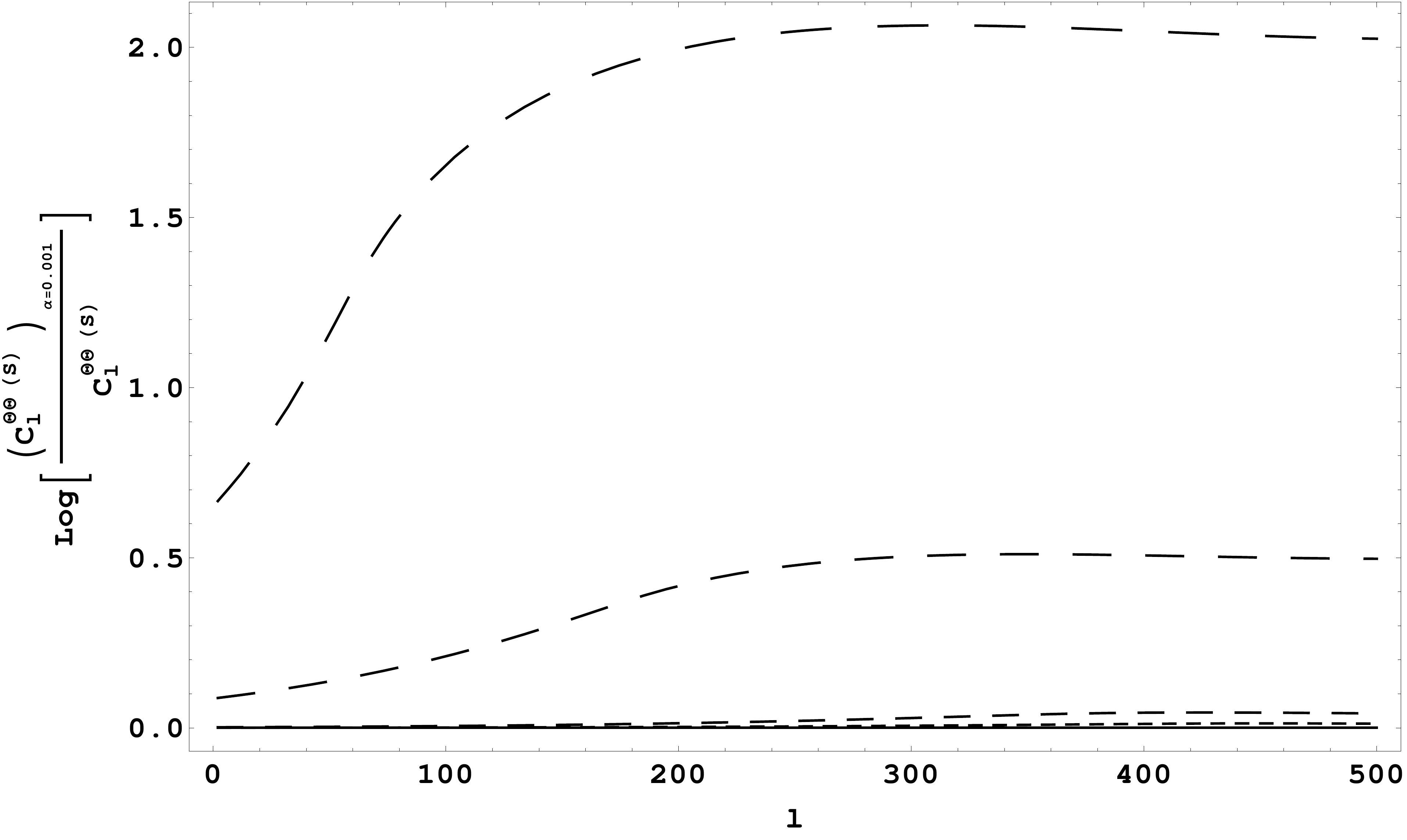}
  \caption{This plot shows again a comparison between the  CMB temperature anisotropy
angular power spectrum  induced by scalar PMF at $k_m=0.001k_D$ lower cutoff,  respect to the other ones with different values of infrared cutoff. Here, the solid horizontal line is for $k_m=0.1k_D$. The dashed lines describe  $k_m=0.3k_D$, $k_m=0.4k_D$,  $k_m=0.7k_D$, and $k_m=0.9k_D$  from the small to the longest
dashed   lines  respectively.}
  \label{clsT2}
\end{figure}
The figures \ref{clsT1} and \ref{clsT2} show the effects of PMF on the  scalar mode of CMB spectra. Here we did a comparison between the Cls with a null cutoff respect to Cls generated by values of cutoff different from zero. The horizontal solid line shows the comparison with $k_m=0$, $k_m=0.001k_D$, $k_m=0.1k_D$;   no difference in effectiveness was found between these values. The dashed lines report a significant difference of the Cls for values of  $k_m=0.3k_D$,  $k_m=0.7k_D$, and  $k_m=0.9k_D$. In  figure   \ref{clsT2.5}  we show  the dependence of the anisotropic trace-free tensor part power spectrum with the infrared cutoff. We observe again a strong dependence for values larger than $0.2k_D$  represented by the dashed line.   In fact, from figures  \ref{clsT3} and \ref{clsT4}, we find that tensor modes of the CMB spectra are distorted by values of $\alpha$ greater than 0.2. 
It is appropiate to remark that power spectrum of causal fields is a smooth function in the k-space without any sharp cutoff coming from the original mechanism, now,  
given the parametrization introduced in this paper we notice from figure \ref{comparison} that for $\alpha$ very small, the calculations agree with previous
work. It can be thinking as  contribution of the super horizon modes is  negligible and one would expect that scales as $\sim k^4$ for instance. But the results found here have demostrated that an infrared 
cutoff plays an important role in physical scenarios in other cases where $\alpha>0.2$.  Also, one of the characteristics of this dependence is the existence of a peak; indeed, for large values of $\alpha$  the peak moves to left as we see for instance with $\alpha=0.4$ where the peak is in $l\sim 380$ while for $\alpha=0.9$ the peak is shifted to    $l\sim 200$.

\begin{figure}[h!]
  \centering
    \includegraphics[width=.48\textwidth]{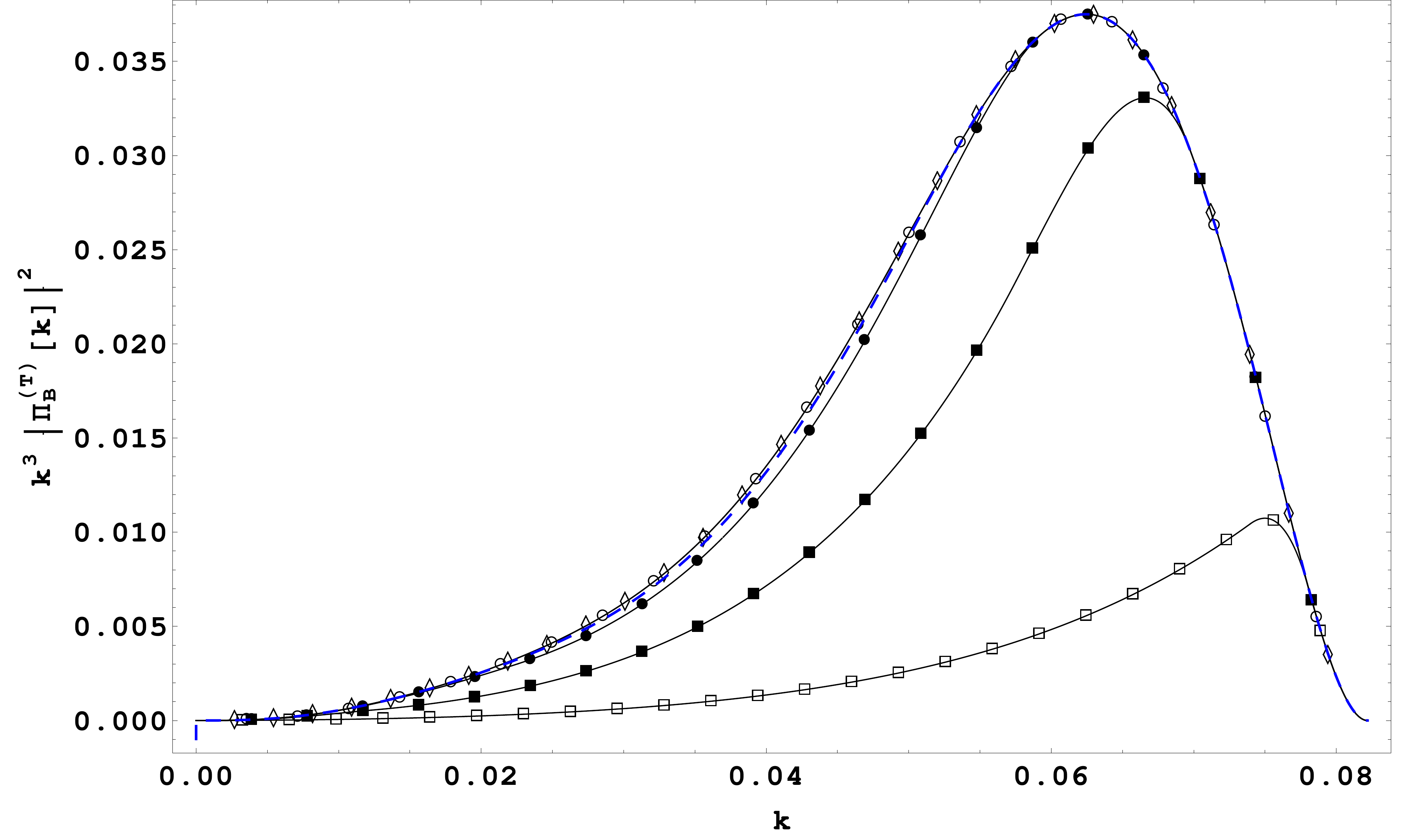}
  \caption{Plot of anisotropic trace-free tensor part power spectrum $k^3\left|\Pi^{(T)}(k,\tau)\right|^2$ versus $k(\mbox{Mpc}^{-1})$ with $n_B=2$ for different values of infrared-cutoff, lines with open squares refer to  $k_m=0.9k_D$. Lines with filled squares correspond to  $k_m=0.7k_D$, lines with filled circles for $k_m=0.5k_D$; dashed line refer to $k_m=0.3k_D$, and finally, lines with open circles and diamonds correspond to $k_m=0.1k_D$ and  $k_m=0.001k_D$ respectively. }
  \label{clsT2.5}
\end{figure}
\begin{figure}[h!]
  \centering
    \includegraphics[width=.48\textwidth]{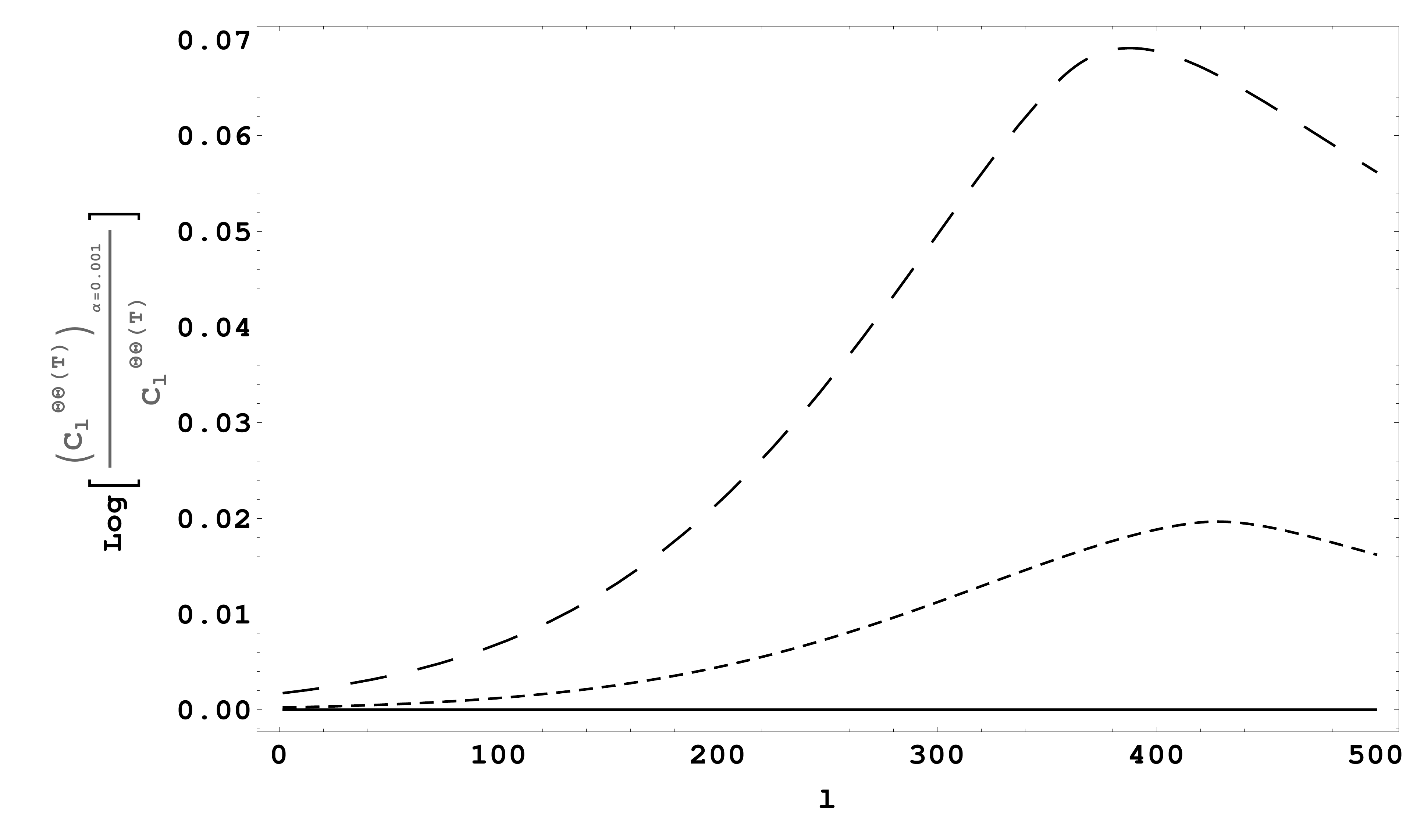}
  \caption{Comparison between the  CMB temperature anisotropy
angular power spectrum  induced by tensor magnetic perturbation  at $k_m=0.001k_D$ lower cutoff,  respect to the other ones with different values of infrared cutoff. Here, the solid horizon line is for $k_m=0.1k_D$; small and large dashed  lines refer to $k_m=0.3k_D$ and  $k_m=0.4k_D$ respectively.  }
  \label{clsT3}
\end{figure}
\begin{figure}[h!]
  \centering
    \includegraphics[width=.48\textwidth]{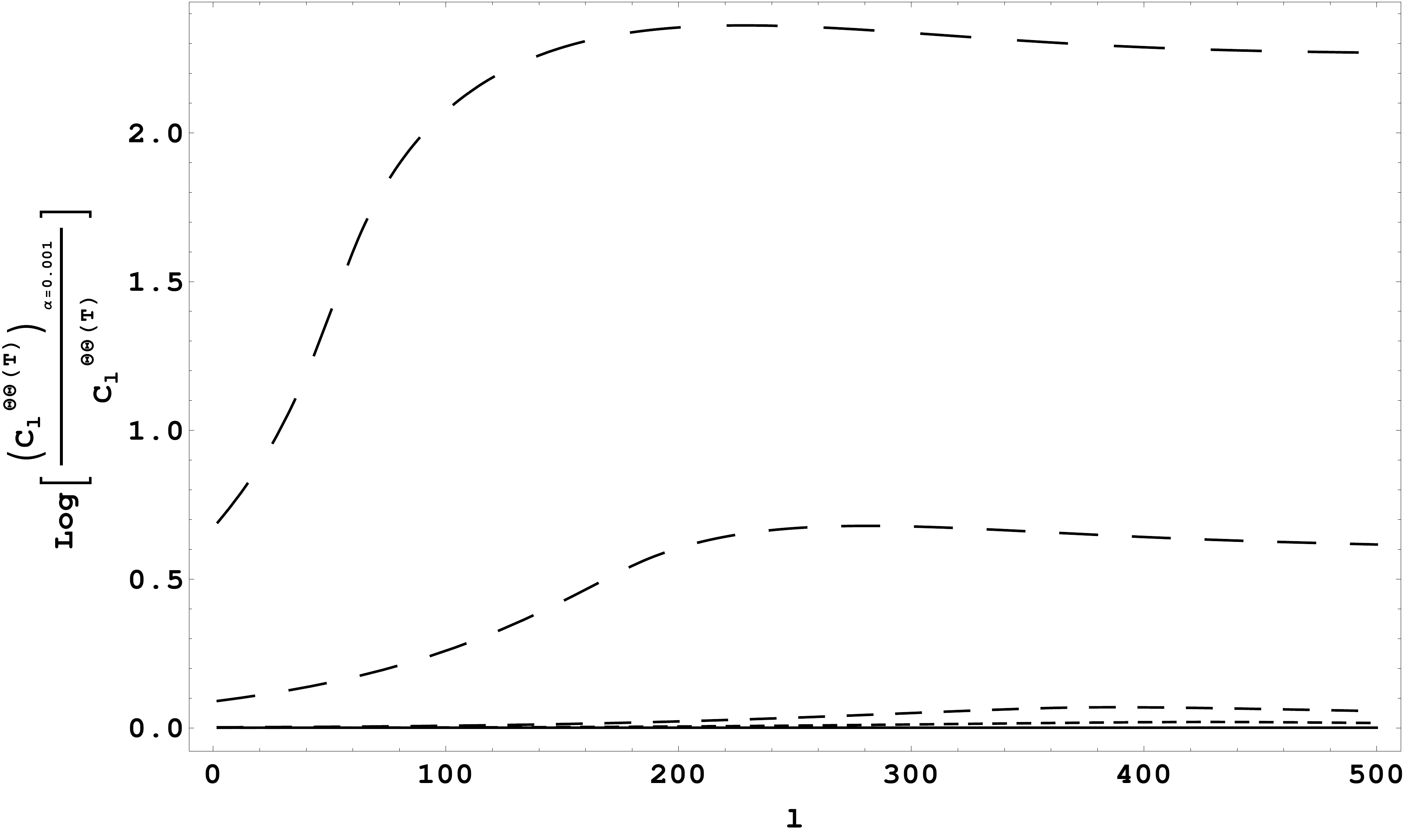}
  \caption{This plot shows again a comparison between the  CMB temperature anisotropy
angular power spectrum  induced by tensor magnetic fluctuations at $k_m=0.001k_D$ lower cutoff,  respect to the other ones with different values of infrared cutoff. Here, the solid horizon line is for $k_m=0.1k_D$. The dashed lines describe  $k_m=0.3k_D$, $k_m=0.4k_D$,  $k_m=0.7k_D$, and $k_m=0.9k_D$  from the small to the longest
dashed   lines  respectively. }
  \label{clsT4}
\end{figure}
In summary  we are working on the assumption that after  inflation  a weak  magnetic field, a seed,  was created. This PMF is parametrized  by its strength $B_\lambda$, smoothing length $\lambda$ and in accordance with the generation process, it also  depends on $k_D$, $k_m$ and a blue spectral index $n_B$. In particular, $k_m$  is set  by the size of the causal part of the Universe during its generation. Now, if this seed indeed is presented during late stages in the universe, this PMF prints a signal  in the pattern on CMB spectra, signal that depends of the variables above mentioned, in particular $k_m$. 
If $\alpha$ is close to one the effect of infrared cutoff must not be ignored, even  in scenarios like inflation this cutoff is also important  (For a deeper discussion see \cite{yamazaki3}).   Therefore,  the feature of this signal which we found is strongly dependent of the infrared cutoff, will be  useful for constraining  PMF post inflation generation models.  Besides this $k_m$ is important for studying the evolution of density perturbations and peculiar velocities due to primordial magnetic fields and effects on BBN  \cite{yamazaki3}, \cite{yamazaki4}, \cite{olinto}, \cite{jedamzikolinto2}.

\section{Discussion}
The origin of large scale magnetic fields is one of the most puzzling topics in cosmology and astrophysics. Understanding its generation and evolution is a main goal from both theoretical and observational aspects.
In this work we have discussed how magnetic fields created in early epochs in the universe: PMFs,  could affect the power spectrum  on CMB pattern temperature.  These  PMFs can be characterized  by  the amplitud of the field  and its spectral index in according to the generation model, supposing a power law scaling. The power spectra for causal  PMFs or  post inflationary fields is strongly dependent of an upper cutoff (due to damped on small scales by radiation viscosity), a lower cutoff determined by the  the causal horizon size, and has the  property  that $n\geq 2$.  Here, we use this insight to  solve the exact  convolution of the Fourier spectra for scalar, vector and tensor modes,  to improve a previously estimation proposed by  \cite{tina1} and \cite{paoletti1}. The main difference lies in the fact that we consider a lower cutoff which  takes into consideration only those modes inside the causal region. We have shown the exact power spectrum for a PMFs choosing a small infrared cutoff and finding a good agreement with  \cite{paoletti1}. Next, we use these results for calculating the angular power spectrum of CMB anisotropies due to a PMF and we get the results shown in figures \ref{cL1} and \ref{clsT} which are  in good concordance with  the obtained by \cite{tina1}. However, in considering just causal fields, the infrared cutoff starts to be relevant in the power spectrum of these fields as we found in figures \ref{variation} and \ref{clsT2.5}  where for values of $\alpha>0.2$ being $\alpha=k_m/k_D$, the PMF spectra changes drastically, except to values $k$ close to $2k_D$ whose slope remains invariant.  We also found that for large values of $\alpha$ the peak of the spectrum moves to high wavenumbers. Hence, if the value of the lower cutoff changes, the CMB spectra would have to be distorted by this change and therefore, observing this effect of CMB we could infer the value of this cutoff and thus constraining PMF post inflation generation model. The dependence of distortion of CMB spectra respect to infrared cutoff  was shown in the figures \ref{clsT1}, \ref{clsT2} for scalar modes and \ref{clsT3}, \ref{clsT4} for tensor modes. 
In conclusion, constrainting the value of $\alpha$ via CMB observations, we offers the possibility to set  the epoch where PMF was created in order to distinguish the cosmological model in which the seed field was  produced.  
\section*{Acknowledgments}

We greatly appreciate useful comments from Kerstin  Kunze and Tina Kahniashvili.
 
\appendix 
\section{Integration domain}
\label{appendix}
The  conditions  over $k$  equation (\ref{cond}), introduce a dependence on the angular integration domain and the two allow the energy
power spectrum to be non zero only for $0 <k<2k_D$. The conditions split the double integral in the following form, for $0.2k_D>k_m>0$ we have
\begin{equation}
\boxed{2k_m>k>0}
\end{equation}
\begin{eqnarray}
&&\int_{k_m}^{k+k_m}d^3 k^\prime \int_{-1}^{\frac{k^2+k^{\prime \, 2}-k_m^2}{2kk^\prime}}d \gamma +\nonumber\\
&&\int_{k_m+k}^{k_D-k}d^3k^\prime \int_{-1}^{1}d \gamma+\int_{k_D-k}^{k_D}d^3k^\prime \int^{1}_{\frac{k^2+k^{\prime\, 2}-k_D^2}{2kk^\prime}}d \gamma\nonumber
\end{eqnarray}
\begin{equation}
\boxed{\frac{k_D-k_m}{2}>k>2k_m}
\end{equation}
\begin{eqnarray}
&&\int_{k_m}^{k-k_m}d^3 k^\prime \int_{-1}^{1}d \gamma+\int_{k_m+k}^{k_D-k}d^3k^\prime \int_{-1}^{1}d \gamma +\nonumber\\
&&\int_{k-k_m}^{k+k_m}d^3k^\prime \int_{-1}^{\frac{k^2+k^{\prime\, 2}-k_m^2}{2kk^\prime}}d \gamma+\int_{k_D-k}^{k_D}d^3k^\prime \int^{1}_{\frac{k^2+k^{\prime\, 2}-k_D^2}{2kk^\prime}}d \gamma\nonumber
\end{eqnarray}

\begin{equation}
\boxed{\frac{k_D+k_m}{2}>k>\frac{k_D-k_m}{2}}
\end{equation}
\begin{eqnarray}
&&\int_{k_m}^{k-k_m}d^3 k^\prime \int_{-1}^{1}d \gamma+\int_{k-k_m}^{k_D-k}d^3k^\prime \int_{-1}^{\frac{k^2+k^{\prime\, 2}-k_m^2}{2kk^\prime}}d \gamma +\nonumber\\
&&\int_{k_D-k}^{k+k_m}d^3k^\prime \int_{\frac{k^2+k^{\prime\, 2}-k_D^2}{2kk^\prime}}^{\frac{k^2+k^{\prime\, 2}-k_m^2}{2kk^\prime}}d \gamma+\int_{k_m+k}^{k_D}d^3k^\prime \int^{1}_{\frac{k^2+k^{\prime\, 2}-k_D^2}{2kk^\prime}}d \gamma \nonumber
\end{eqnarray}

\begin{equation}
\boxed{k_D-k_m>k>\frac{k_D+k_m}{2}}
\end{equation}
\begin{eqnarray}
&&\int_{k_m}^{k_D-k}d^3 k^\prime \int_{-1}^{1}d \gamma+\int_{k_D-k}^{k-k_m}d^3k^\prime \int^{1}_{\frac{k^2+k^{\prime\, 2}-k_D^2}{2kk^\prime}}d \gamma +\nonumber\\
&&\int_{k-k_m}^{k+k_m}d^3k^\prime \int_{\frac{k^2+k^{\prime\, 2}-k_D^2}{2kk^\prime}}^{\frac{k^2+k^{\prime\, 2}-k_m^2}{2kk^\prime}}d \gamma+\int_{k_m+k}^{k_D}d^3k^\prime \int^{1}_{\frac{k^2+k^{\prime\, 2}-k_D^2}{2kk^\prime}}d \gamma \nonumber
\end{eqnarray}

\begin{equation}
\boxed{k_D+k_m>k>k_D-k_m}
\end{equation}
\begin{eqnarray}
\int_{k_m}^{k-k_m}d^3 k^\prime \int_{\frac{k^2+k^{\prime\, 2}-k_D^2}{2kk^\prime}}^{1}d \gamma&+&\int_{k-k_m}^{k_D}d^3k^\prime \int^{\frac{k^2+k^{\prime\, 2}-k_m^2}{2kk^\prime}}_{\frac{k^2+k^{\prime\, 2}-k_D^2}{2kk^\prime}}d \gamma  \nonumber
\end{eqnarray}

\begin{equation}
\boxed{2k_D>k>k_D+k_m}
\end{equation}
\begin{equation}
\int_{k-k_D}^{k_D}d^3 k^\prime \int_{\frac{k^2+k^{\prime\, 2}-k_D^2}{2kk^\prime}}^{1}d \gamma. \nonumber
\end{equation}

For the case where $k_D>k_m>0.2k_D$,  we have
\begin{equation}
\boxed{\frac{k_D-k_m}{2}>k>0}
\end{equation}
\begin{eqnarray}
&&\int_{k_m}^{k+k_m}d^3 k^\prime \int_{-1}^{\frac{k^2+k^{\prime \, 2}-k_m^2}{2kk^\prime}}d \gamma+\int_{k_m+k}^{k_D-k}d^3k^\prime \int_{-1}^{1}d \gamma+\nonumber\\
&&\int_{k_D-k}^{k_D}d^3k^\prime \int^{1}_{\frac{k^2+k^{\prime\, 2}-k_D^2}{2kk^\prime}}d \gamma \nonumber
\end{eqnarray}

\begin{equation}
\boxed{k_D-k_m>k>\frac{k_D-k_m}{2}}
\end{equation}

\begin{eqnarray}
&&\int_{k_m}^{k-k_m}d^3 k^\prime \int_{-1}^{\frac{k^2+k^{\prime\, 2}-k_m^2}{2kk^\prime}}d \gamma+\int_{k_D-k}^{k_m+k}d^3k^\prime \int_{\frac{k^2+k^{\prime\, 2}-k_D^2}{2kk^\prime}}^{\frac{k^2+k^{\prime\, 2}-k_m^2}{2kk^\prime}}d \gamma +\nonumber \\
&&\int_{k+k_m}^{k_D}d^3k^\prime \int^{1}_{\frac{k^2+k^{\prime\, 2}-k_D^2}{2kk^\prime}}d \gamma \nonumber
\end{eqnarray}

\begin{equation}
\boxed{2k_m>k>k_D-k_m}
\end{equation}

\begin{equation}
\int_{k_m}^{k_D}d^3k^\prime \int_{\frac{k^2+k^{\prime\, 2}-k_D^2}{2kk^\prime}}^{\frac{k^2+k^{\prime\, 2}-k_m^2}{2kk^\prime}}d \gamma \nonumber
\end{equation}

\begin{equation}
\boxed{k_m+k_D>k>2k_m}
\end{equation}

\begin{equation}
\int_{k-k_m}^{k_D}d^3k^\prime \int_{\frac{k^2+k^{\prime\, 2}-k_D^2}{2kk^\prime}}^{\frac{k^2+k^{\prime\, 2}-k_m^2}{2kk^\prime}}d \gamma+\int_{k_m}^{k-k_m}d^3k^\prime \int^{1}_{\frac{k^2+k^{\prime\, 2}-k_D^2}{2kk^\prime}}d \gamma \nonumber
\end{equation}
\begin{equation}
\boxed{2k_D>k>k_m+k_D}
\end{equation}
\begin{equation}
\int_{k-k_D}^{k_D}d^3k^\prime \int^{1}_{\frac{k^2+k^{\prime\, 2}-k_D^2}{2kk^\prime}}d \gamma. \nonumber
\end{equation}



In the case where $k_m=0$, the integration domain  leads to

\begin{equation}
\boxed{k_D>k>0}
\end{equation}
\begin{equation}
\int_{0}^{k_D-k}d^3 k^\prime \int_{-1}^{1}d \gamma+\int_{k_D-k}^{k_D}d^3k^\prime \int_{\frac{k^2+k^{\prime\, 2}-k_D^2}{2kk^\prime}}^{1}d \gamma \nonumber
\end{equation}
\begin{equation}
\boxed{2k_D>k>k_D}
\end{equation}
\begin{equation}
\int_{k-k_D}^{k_D}d^3k^\prime \int_{\frac{k^2+k^{\prime\, 2}-k_D^2}{2kk^\prime}}^{1}d \gamma, \nonumber
\end{equation}
which is in agreement with \cite{paoletti1}.
\bibliographystyle{apsrev4-1}
\bibliography{apssamp}

\begin{thebibliography}{38}%
\makeatletter
\providecommand \@ifxundefined [1]{%
 \@ifx{#1\undefined}
}%
\providecommand \@ifnum [1]{%
 \ifnum #1\expandafter \@firstoftwo
 \else \expandafter \@secondoftwo
 \fi
}%
\providecommand \@ifx [1]{%
 \ifx #1\expandafter \@firstoftwo
 \else \expandafter \@secondoftwo
 \fi
}%
\providecommand \natexlab [1]{#1}%
\providecommand \enquote  [1]{``#1''}%
\providecommand \bibnamefont  [1]{#1}%
\providecommand \bibfnamefont [1]{#1}%
\providecommand \citenamefont [1]{#1}%
\providecommand \href@noop [0]{\@secondoftwo}%
\providecommand \href [0]{\begingroup \@sanitize@url \@href}%
\providecommand \@href[1]{\@@startlink{#1}\@@href}%
\providecommand \@@href[1]{\endgroup#1\@@endlink}%
\providecommand \@sanitize@url [0]{\catcode `\\12\catcode `\$12\catcode
  `\&12\catcode `\#12\catcode `\^12\catcode `\_12\catcode `\%12\relax}%
\providecommand \@@startlink[1]{}%
\providecommand \@@endlink[0]{}%
\providecommand \url  [0]{\begingroup\@sanitize@url \@url }%
\providecommand \@url [1]{\endgroup\@href {#1}{\urlprefix }}%
\providecommand \urlprefix  [0]{URL }%
\providecommand \Eprint [0]{\href }%
\providecommand \doibase [0]{http://dx.doi.org/}%
\providecommand \selectlanguage [0]{\@gobble}%
\providecommand \bibinfo  [0]{\@secondoftwo}%
\providecommand \bibfield  [0]{\@secondoftwo}%
\providecommand \translation [1]{[#1]}%
\providecommand \BibitemOpen [0]{}%
\providecommand \bibitemStop [0]{}%
\providecommand \bibitemNoStop [0]{.\EOS\space}%
\providecommand \EOS [0]{\spacefactor3000\relax}%
\providecommand \BibitemShut  [1]{\csname bibitem#1\endcsname}%
\let\auto@bib@innerbib\@empty
\bibitem [{\citenamefont {{Widrow}}(2002)}]{Lawrence}%
  \BibitemOpen
  \bibfield  {author} {\bibinfo {author} {\bibfnamefont {L.~M.}\ \bibnamefont
  {{Widrow}}},\ }\href {\doibase 10.1103/RevModPhys.74.775} {\bibfield
  {journal} {\bibinfo  {journal} {Reviews of Modern Physics}\ }\textbf
  {\bibinfo {volume} {74}},\ \bibinfo {pages} {775} (\bibinfo {year} {2002})},\
  \Eprint {http://arxiv.org/abs/astro-ph/0207240} {astro-ph/0207240}
  \BibitemShut {NoStop}%
\bibitem [{\citenamefont {{Neronov}}\ and\ \citenamefont
  {{Vovk}}(2010)}]{neronov}%
  \BibitemOpen
  \bibfield  {author} {\bibinfo {author} {\bibfnamefont {A.}~\bibnamefont
  {{Neronov}}}\ and\ \bibinfo {author} {\bibfnamefont {I.}~\bibnamefont
  {{Vovk}}},\ }\href {\doibase 10.1126/science.1184192} {\bibfield  {journal}
  {\bibinfo  {journal} {Science}\ }\textbf {\bibinfo {volume} {328}},\ \bibinfo
  {pages} {73} (\bibinfo {year} {2010})},\ \Eprint
  {http://arxiv.org/abs/1006.3504} {arXiv:1006.3504 [astro-ph.HE]} \BibitemShut
  {NoStop}%
\bibitem [{\citenamefont {{Kandus}}\ \emph {et~al.}(2011)\citenamefont
  {{Kandus}}, \citenamefont {{Kunze}},\ and\ \citenamefont
  {{Tsagas}}}]{Kandus}%
  \BibitemOpen
  \bibfield  {author} {\bibinfo {author} {\bibfnamefont {A.}~\bibnamefont
  {{Kandus}}}, \bibinfo {author} {\bibfnamefont {K.~E.}\ \bibnamefont
  {{Kunze}}}, \ and\ \bibinfo {author} {\bibfnamefont {C.~G.}\ \bibnamefont
  {{Tsagas}}},\ }\href {\doibase 10.1016/j.physrep.2011.03.001} {\bibfield
  {journal} {\bibinfo  {journal} {Phys.Rept}\ }\textbf {\bibinfo {volume}
  {505}},\ \bibinfo {pages} {1} (\bibinfo {year} {2011})},\ \Eprint
  {http://arxiv.org/abs/1007.3891} {arXiv:1007.3891 [astro-ph.CO]} \BibitemShut
  {NoStop}%
\bibitem [{\citenamefont {{Banerjee}}\ and\ \citenamefont
  {{Jedamzik}}(2003)}]{banerjee1}%
  \BibitemOpen
  \bibfield  {author} {\bibinfo {author} {\bibfnamefont {R.}~\bibnamefont
  {{Banerjee}}}\ and\ \bibinfo {author} {\bibfnamefont {K.}~\bibnamefont
  {{Jedamzik}}},\ }\href {\doibase 10.1103/PhysRevLett.91.251301} {\bibfield
  {journal} {\bibinfo  {journal} {Physical Review Letters}\ }\textbf {\bibinfo
  {volume} {91}},\ \bibinfo {eid} {251301} (\bibinfo {year} {2003})},\ \Eprint
  {http://arxiv.org/abs/astro-ph/0306211} {astro-ph/0306211} \BibitemShut
  {NoStop}%
\bibitem [{\citenamefont {{Subramanian}}\ and\ \citenamefont
  {{Barrow}}(1998)}]{barrow}%
  \BibitemOpen
  \bibfield  {author} {\bibinfo {author} {\bibfnamefont {K.}~\bibnamefont
  {{Subramanian}}}\ and\ \bibinfo {author} {\bibfnamefont {J.~D.}\ \bibnamefont
  {{Barrow}}},\ }\href {\doibase 10.1103/PhysRevD.58.083502} {\bibfield
  {journal} {\bibinfo  {journal} {\prd}\ }\textbf {\bibinfo {volume} {58}},\
  \bibinfo {eid} {083502} (\bibinfo {year} {1998})},\ \Eprint
  {http://arxiv.org/abs/astro-ph/9712083} {astro-ph/9712083} \BibitemShut
  {NoStop}%
\bibitem [{\citenamefont {{Jedamzik}}\ and\ \citenamefont
  {{Sigl}}(2011)}]{sigl2}%
  \BibitemOpen
  \bibfield  {author} {\bibinfo {author} {\bibfnamefont {K.}~\bibnamefont
  {{Jedamzik}}}\ and\ \bibinfo {author} {\bibfnamefont {G.}~\bibnamefont
  {{Sigl}}},\ }\href {\doibase 10.1103/PhysRevD.83.103005} {\bibfield
  {journal} {\bibinfo  {journal} {\prd}\ }\textbf {\bibinfo {volume} {83}},\
  \bibinfo {eid} {103005} (\bibinfo {year} {2011})},\ \Eprint
  {http://arxiv.org/abs/1012.4794} {arXiv:1012.4794 [astro-ph.CO]} \BibitemShut
  {NoStop}%
\bibitem [{\citenamefont {{Banerjee}}\ and\ \citenamefont
  {{Jedamzik}}(2004)}]{banerjee}%
  \BibitemOpen
  \bibfield  {author} {\bibinfo {author} {\bibfnamefont {R.}~\bibnamefont
  {{Banerjee}}}\ and\ \bibinfo {author} {\bibfnamefont {K.}~\bibnamefont
  {{Jedamzik}}},\ }\href {\doibase 10.1103/PhysRevD.70.123003} {\bibfield
  {journal} {\bibinfo  {journal} {\prd}\ }\textbf {\bibinfo {volume} {70}},\
  \bibinfo {eid} {123003} (\bibinfo {year} {2004})},\ \Eprint
  {http://arxiv.org/abs/astro-ph/0410032} {astro-ph/0410032} \BibitemShut
  {NoStop}%
\bibitem [{\citenamefont {{Saveliev}}\ \emph {et~al.}(2013)\citenamefont
  {{Saveliev}}, \citenamefont {{Jedamzik}},\ and\ \citenamefont
  {{Sigl}}}]{sigl1}%
  \BibitemOpen
  \bibfield  {author} {\bibinfo {author} {\bibfnamefont {A.}~\bibnamefont
  {{Saveliev}}}, \bibinfo {author} {\bibfnamefont {K.}~\bibnamefont
  {{Jedamzik}}}, \ and\ \bibinfo {author} {\bibfnamefont {G.}~\bibnamefont
  {{Sigl}}},\ }\href {\doibase 10.1103/PhysRevD.87.123001} {\bibfield
  {journal} {\bibinfo  {journal} {\prd}\ }\textbf {\bibinfo {volume} {87}},\
  \bibinfo {eid} {123001} (\bibinfo {year} {2013})},\ \Eprint
  {http://arxiv.org/abs/1304.3621} {arXiv:1304.3621 [astro-ph.CO]} \BibitemShut
  {NoStop}%
\bibitem [{\citenamefont {{Giovannini}}(2008)}]{giovannini2}%
  \BibitemOpen
  \bibfield  {author} {\bibinfo {author} {\bibfnamefont {M.}~\bibnamefont
  {{Giovannini}}},\ }in\ \href@noop {} {\emph {\bibinfo {booktitle} {String
  Theory and Fundamental Interactions}}},\ \bibinfo {series} {Lecture Notes in
  Physics, Berlin Springer Verlag}, Vol.\ \bibinfo {volume} {737},\ \bibinfo
  {editor} {edited by\ \bibinfo {editor} {\bibfnamefont {M.}~\bibnamefont
  {{Gasperini}}}\ and\ \bibinfo {editor} {\bibfnamefont {J.}~\bibnamefont
  {{Maharana}}}}\ (\bibinfo {year} {2008})\ p.\ \bibinfo {pages} {863},\
  \Eprint {http://arxiv.org/abs/astro-ph/0612378} {astro-ph/0612378}
  \BibitemShut {NoStop}%
\bibitem [{\citenamefont {Choudhury}(2014)}]{sayantan}%
  \BibitemOpen
  \bibfield  {author} {\bibinfo {author} {\bibfnamefont {S.}~\bibnamefont
  {Choudhury}},\ }\href {\doibase
  http://dx.doi.org/10.1016/j.physletb.2014.06.029} {\bibfield  {journal}
  {\bibinfo  {journal} {Physics Letters B}\ }\textbf {\bibinfo {volume}
  {735}},\ \bibinfo {pages} {138 } (\bibinfo {year} {2014})}\BibitemShut
  {NoStop}%
\bibitem [{\citenamefont {{Mack}}\ \emph {et~al.}(2002)\citenamefont {{Mack}},
  \citenamefont {{Kahniashvili}},\ and\ \citenamefont {{Kosowsky}}}]{tina1}%
  \BibitemOpen
  \bibfield  {author} {\bibinfo {author} {\bibfnamefont {A.}~\bibnamefont
  {{Mack}}}, \bibinfo {author} {\bibfnamefont {T.}~\bibnamefont
  {{Kahniashvili}}}, \ and\ \bibinfo {author} {\bibfnamefont {A.}~\bibnamefont
  {{Kosowsky}}},\ }\href {\doibase 10.1103/PhysRevD.65.123004} {\bibfield
  {journal} {\bibinfo  {journal} {\prd}\ }\textbf {\bibinfo {volume} {65}},\
  \bibinfo {eid} {123004} (\bibinfo {year} {2002})},\ \Eprint
  {http://arxiv.org/abs/astro-ph/0105504} {astro-ph/0105504} \BibitemShut
  {NoStop}%
\bibitem [{\citenamefont {{Grasso}}\ and\ \citenamefont
  {{Rubinstein}}(2001)}]{Grasso}%
  \BibitemOpen
  \bibfield  {author} {\bibinfo {author} {\bibfnamefont {D.}~\bibnamefont
  {{Grasso}}}\ and\ \bibinfo {author} {\bibfnamefont {H.~R.}\ \bibnamefont
  {{Rubinstein}}},\ }\href {\doibase 10.1016/S0370-1573(00)00110-1} {\bibfield
  {journal} {\bibinfo  {journal} {Phys.Rept}\ }\textbf {\bibinfo {volume}
  {348}},\ \bibinfo {pages} {163} (\bibinfo {year} {2001})},\ \Eprint
  {http://arxiv.org/abs/astro-ph/0009061} {astro-ph/0009061} \BibitemShut
  {NoStop}%
\bibitem [{\citenamefont {{Giovannini}}\ and\ \citenamefont
  {{Kunze}}(2008)}]{giovannini-kunze}%
  \BibitemOpen
  \bibfield  {author} {\bibinfo {author} {\bibfnamefont {M.}~\bibnamefont
  {{Giovannini}}}\ and\ \bibinfo {author} {\bibfnamefont {K.~E.}\ \bibnamefont
  {{Kunze}}},\ }\href {\doibase 10.1103/PhysRevD.77.063003} {\bibfield
  {journal} {\bibinfo  {journal} {\prd}\ }\textbf {\bibinfo {volume} {77}},\
  \bibinfo {eid} {063003} (\bibinfo {year} {2008})},\ \Eprint
  {http://arxiv.org/abs/0712.3483} {arXiv:0712.3483} \BibitemShut {NoStop}%
\bibitem [{\citenamefont {{Kunze}}(2014)}]{kunze2}%
  \BibitemOpen
  \bibfield  {author} {\bibinfo {author} {\bibfnamefont {K.~E.}\ \bibnamefont
  {{Kunze}}},\ }\href {\doibase 10.1103/PhysRevD.89.103016} {\bibfield
  {journal} {\bibinfo  {journal} {\prd}\ }\textbf {\bibinfo {volume} {89}},\
  \bibinfo {eid} {103016} (\bibinfo {year} {2014})},\ \Eprint
  {http://arxiv.org/abs/1312.5630} {arXiv:1312.5630} \BibitemShut {NoStop}%
\bibitem [{\citenamefont {{Giovannini}}(2004)}]{giovannini1}%
  \BibitemOpen
  \bibfield  {author} {\bibinfo {author} {\bibfnamefont {M.}~\bibnamefont
  {{Giovannini}}},\ }\href {\doibase 10.1103/PhysRevD.70.123507} {\bibfield
  {journal} {\bibinfo  {journal} {\prd}\ }\textbf {\bibinfo {volume} {70}},\
  \bibinfo {eid} {123507} (\bibinfo {year} {2004})},\ \Eprint
  {http://arxiv.org/abs/astro-ph/0409594} {astro-ph/0409594} \BibitemShut
  {NoStop}%
\bibitem [{\citenamefont {{Shaw}}\ and\ \citenamefont {{Lewis}}(2010)}]{lewis}%
  \BibitemOpen
  \bibfield  {author} {\bibinfo {author} {\bibfnamefont {J.~R.}\ \bibnamefont
  {{Shaw}}}\ and\ \bibinfo {author} {\bibfnamefont {A.}~\bibnamefont
  {{Lewis}}},\ }\href {\doibase 10.1103/PhysRevD.81.043517} {\bibfield
  {journal} {\bibinfo  {journal} {\prd}\ }\textbf {\bibinfo {volume} {81}},\
  \bibinfo {eid} {043517} (\bibinfo {year} {2010})},\ \Eprint
  {http://arxiv.org/abs/0911.2714} {arXiv:0911.2714 [astro-ph.CO]} \BibitemShut
  {NoStop}%
\bibitem [{\citenamefont {{Kunze}}(2011)}]{kunze1}%
  \BibitemOpen
  \bibfield  {author} {\bibinfo {author} {\bibfnamefont {K.~E.}\ \bibnamefont
  {{Kunze}}},\ }\href {\doibase 10.1103/PhysRevD.83.023006} {\bibfield
  {journal} {\bibinfo  {journal} {\prd}\ }\textbf {\bibinfo {volume} {83}},\
  \bibinfo {eid} {023006} (\bibinfo {year} {2011})},\ \Eprint
  {http://arxiv.org/abs/1007.3163} {arXiv:1007.3163 [astro-ph.CO]} \BibitemShut
  {NoStop}%
\bibitem [{\citenamefont {{Yamazaki}}\ \emph {et~al.}(2008)\citenamefont
  {{Yamazaki}}, \citenamefont {{Ichiki}}, \citenamefont {{Kajino}},\ and\
  \citenamefont {{Mathews}}}]{yamazaki1}%
  \BibitemOpen
  \bibfield  {author} {\bibinfo {author} {\bibfnamefont {D.~G.}\ \bibnamefont
  {{Yamazaki}}}, \bibinfo {author} {\bibfnamefont {K.}~\bibnamefont
  {{Ichiki}}}, \bibinfo {author} {\bibfnamefont {T.}~\bibnamefont {{Kajino}}},
  \ and\ \bibinfo {author} {\bibfnamefont {G.~J.}\ \bibnamefont {{Mathews}}},\
  }\href {\doibase 10.1103/PhysRevD.77.043005} {\bibfield  {journal} {\bibinfo
  {journal} {\prd}\ }\textbf {\bibinfo {volume} {77}},\ \bibinfo {eid} {043005}
  (\bibinfo {year} {2008})},\ \Eprint {http://arxiv.org/abs/0801.2572}
  {arXiv:0801.2572} \BibitemShut {NoStop}%
\bibitem [{\citenamefont {{Durrer}}(2007)}]{durrer1}%
  \BibitemOpen
  \bibfield  {author} {\bibinfo {author} {\bibfnamefont {R.}~\bibnamefont
  {{Durrer}}},\ }\href {\doibase 10.1016/j.newar.2006.11.057} {\bibfield
  {journal} {\bibinfo  {journal} {NewAstron.Rev}\ }\textbf {\bibinfo {volume}
  {51}},\ \bibinfo {pages} {275} (\bibinfo {year} {2007})},\ \Eprint
  {http://arxiv.org/abs/astro-ph/0609216} {astro-ph/0609216} \BibitemShut
  {NoStop}%
\bibitem [{\citenamefont {{Durrer}}\ \emph {et~al.}(1998)\citenamefont
  {{Durrer}}, \citenamefont {{Kahniashvili}},\ and\ \citenamefont
  {{Yates}}}]{tina2}%
  \BibitemOpen
  \bibfield  {author} {\bibinfo {author} {\bibfnamefont {R.}~\bibnamefont
  {{Durrer}}}, \bibinfo {author} {\bibfnamefont {T.}~\bibnamefont
  {{Kahniashvili}}}, \ and\ \bibinfo {author} {\bibfnamefont {A.}~\bibnamefont
  {{Yates}}},\ }\href {\doibase 10.1103/PhysRevD.58.123004} {\bibfield
  {journal} {\bibinfo  {journal} {\prd}\ }\textbf {\bibinfo {volume} {58}},\
  \bibinfo {eid} {123004} (\bibinfo {year} {1998})},\ \Eprint
  {http://arxiv.org/abs/astro-ph/9807089} {astro-ph/9807089} \BibitemShut
  {NoStop}%
\bibitem [{\citenamefont {{Durrer}}\ \emph {et~al.}(2000)\citenamefont
  {{Durrer}}, \citenamefont {{Ferreira}},\ and\ \citenamefont
  {{Kahniashvili}}}]{tina3}%
  \BibitemOpen
  \bibfield  {author} {\bibinfo {author} {\bibfnamefont {R.}~\bibnamefont
  {{Durrer}}}, \bibinfo {author} {\bibfnamefont {P.~G.}\ \bibnamefont
  {{Ferreira}}}, \ and\ \bibinfo {author} {\bibfnamefont {T.}~\bibnamefont
  {{Kahniashvili}}},\ }\href {\doibase 10.1103/PhysRevD.61.043001} {\bibfield
  {journal} {\bibinfo  {journal} {\prd}\ }\textbf {\bibinfo {volume} {61}},\
  \bibinfo {eid} {043001} (\bibinfo {year} {2000})},\ \Eprint
  {http://arxiv.org/abs/astro-ph/9911040} {astro-ph/9911040} \BibitemShut
  {NoStop}%
\bibitem [{\citenamefont {{Dreher}}\ \emph {et~al.}(1987)\citenamefont
  {{Dreher}}, \citenamefont {{Carilli}},\ and\ \citenamefont
  {{Perley}}}]{Dreher}%
  \BibitemOpen
  \bibfield  {author} {\bibinfo {author} {\bibfnamefont {J.~W.}\ \bibnamefont
  {{Dreher}}}, \bibinfo {author} {\bibfnamefont {C.~L.}\ \bibnamefont
  {{Carilli}}}, \ and\ \bibinfo {author} {\bibfnamefont {R.~A.}\ \bibnamefont
  {{Perley}}},\ }\href {\doibase 10.1086/165229} {\bibfield  {journal}
  {\bibinfo  {journal} {\apj}\ }\textbf {\bibinfo {volume} {316}},\ \bibinfo
  {pages} {611} (\bibinfo {year} {1987})}\BibitemShut {NoStop}%
\bibitem [{\citenamefont {{Paoletti}}\ \emph {et~al.}(2009)\citenamefont
  {{Paoletti}}, \citenamefont {{Finelli}},\ and\ \citenamefont
  {{Paci}}}]{paoletti}%
  \BibitemOpen
  \bibfield  {author} {\bibinfo {author} {\bibfnamefont {D.}~\bibnamefont
  {{Paoletti}}}, \bibinfo {author} {\bibfnamefont {F.}~\bibnamefont
  {{Finelli}}}, \ and\ \bibinfo {author} {\bibfnamefont {F.}~\bibnamefont
  {{Paci}}},\ }\href {\doibase 10.1111/j.1365-2966.2009.14727.x} {\bibfield
  {journal} {\bibinfo  {journal} {MNRAS}\ }\textbf {\bibinfo {volume} {396}},\
  \bibinfo {pages} {523} (\bibinfo {year} {2009})},\ \Eprint
  {http://arxiv.org/abs/0811.0230} {arXiv:0811.0230} \BibitemShut {NoStop}%
\bibitem [{\citenamefont {{Caprini}}\ \emph {et~al.}(2009)\citenamefont
  {{Caprini}}, \citenamefont {{Finelli}}, \citenamefont {{Paoletti}},\ and\
  \citenamefont {{Riotto}}}]{riotto}%
  \BibitemOpen
  \bibfield  {author} {\bibinfo {author} {\bibfnamefont {C.}~\bibnamefont
  {{Caprini}}}, \bibinfo {author} {\bibfnamefont {F.}~\bibnamefont
  {{Finelli}}}, \bibinfo {author} {\bibfnamefont {D.}~\bibnamefont
  {{Paoletti}}}, \ and\ \bibinfo {author} {\bibfnamefont {A.}~\bibnamefont
  {{Riotto}}},\ }\href {\doibase 10.1088/1475-7516/2009/06/021} {\bibfield
  {journal} {\bibinfo  {journal} {JCAP}\ }\textbf {\bibinfo {volume} {6}},\
  \bibinfo {eid} {021} (\bibinfo {year} {2009})},\ \Eprint
  {http://arxiv.org/abs/0903.1420} {arXiv:0903.1420 [astro-ph.CO]} \BibitemShut
  {NoStop}%
\bibitem [{\citenamefont {{Trivedi}}\ \emph {et~al.}(2014)\citenamefont
  {{Trivedi}}, \citenamefont {{Subramanian}},\ and\ \citenamefont
  {{Seshadri}}}]{trivedi}%
  \BibitemOpen
  \bibfield  {author} {\bibinfo {author} {\bibfnamefont {P.}~\bibnamefont
  {{Trivedi}}}, \bibinfo {author} {\bibfnamefont {K.}~\bibnamefont
  {{Subramanian}}}, \ and\ \bibinfo {author} {\bibfnamefont {T.~R.}\
  \bibnamefont {{Seshadri}}},\ }\href {\doibase 10.1103/PhysRevD.89.043523}
  {\bibfield  {journal} {\bibinfo  {journal} {\prd}\ }\textbf {\bibinfo
  {volume} {89}},\ \bibinfo {eid} {043523} (\bibinfo {year} {2014})},\ \Eprint
  {http://arxiv.org/abs/1312.5308} {arXiv:1312.5308 [astro-ph.CO]} \BibitemShut
  {NoStop}%
\bibitem [{\citenamefont {{Hortua}}\ \emph {et~al.}(2013)\citenamefont
  {{Hortua}}, \citenamefont {{Casta{\~n}eda}},\ and\ \citenamefont
  {{Tejeiro}}}]{hortua}%
  \BibitemOpen
  \bibfield  {author} {\bibinfo {author} {\bibfnamefont {H.~J.}\ \bibnamefont
  {{Hortua}}}, \bibinfo {author} {\bibfnamefont {L.}~\bibnamefont
  {{Casta{\~n}eda}}}, \ and\ \bibinfo {author} {\bibfnamefont {J.~M.}\
  \bibnamefont {{Tejeiro}}},\ }\href {\doibase 10.1103/PhysRevD.87.103531}
  {\bibfield  {journal} {\bibinfo  {journal} {\prd}\ }\textbf {\bibinfo
  {volume} {87}},\ \bibinfo {eid} {103531} (\bibinfo {year} {2013})},\ \Eprint
  {http://arxiv.org/abs/1104.0701} {arXiv:1104.0701 [astro-ph.CO]} \BibitemShut
  {NoStop}%
\bibitem [{\citenamefont {{Durrer}}\ and\ \citenamefont
  {{Kunz}}(1998)}]{kunze}%
  \BibitemOpen
  \bibfield  {author} {\bibinfo {author} {\bibfnamefont {R.}~\bibnamefont
  {{Durrer}}}\ and\ \bibinfo {author} {\bibfnamefont {M.}~\bibnamefont
  {{Kunz}}},\ }\href {\doibase 10.1103/PhysRevD.57.R3199} {\bibfield  {journal}
  {\bibinfo  {journal} {\prd}\ }\textbf {\bibinfo {volume} {57}},\ \bibinfo
  {pages} {3199} (\bibinfo {year} {1998})},\ \Eprint
  {http://arxiv.org/abs/astro-ph/9711133} {astro-ph/9711133} \BibitemShut
  {NoStop}%
\bibitem [{\citenamefont {{Kahniashvili}}\ and\ \citenamefont
  {{Ratra}}(2007)}]{tina4}%
  \BibitemOpen
  \bibfield  {author} {\bibinfo {author} {\bibfnamefont {T.}~\bibnamefont
  {{Kahniashvili}}}\ and\ \bibinfo {author} {\bibfnamefont {B.}~\bibnamefont
  {{Ratra}}},\ }\href {\doibase 10.1103/PhysRevD.75.023002} {\bibfield
  {journal} {\bibinfo  {journal} {\prd}\ }\textbf {\bibinfo {volume} {75}},\
  \bibinfo {eid} {023002} (\bibinfo {year} {2007})},\ \Eprint
  {http://arxiv.org/abs/astro-ph/0611247} {astro-ph/0611247} \BibitemShut
  {NoStop}%
\bibitem [{\citenamefont {{Finelli}}\ \emph {et~al.}(2008)\citenamefont
  {{Finelli}}, \citenamefont {{Paci}},\ and\ \citenamefont
  {{Paoletti}}}]{paoletti1}%
  \BibitemOpen
  \bibfield  {author} {\bibinfo {author} {\bibfnamefont {F.}~\bibnamefont
  {{Finelli}}}, \bibinfo {author} {\bibfnamefont {F.}~\bibnamefont {{Paci}}}, \
  and\ \bibinfo {author} {\bibfnamefont {D.}~\bibnamefont {{Paoletti}}},\
  }\href {\doibase 10.1103/PhysRevD.78.023510} {\bibfield  {journal} {\bibinfo
  {journal} {\prd}\ }\textbf {\bibinfo {volume} {78}},\ \bibinfo {eid} {023510}
  (\bibinfo {year} {2008})},\ \Eprint {http://arxiv.org/abs/0803.1246}
  {arXiv:0803.1246} \BibitemShut {NoStop}%
\bibitem [{\citenamefont {{Jedamzik}}\ \emph {et~al.}(1998)\citenamefont
  {{Jedamzik}}, \citenamefont {{Katalini{\'c}}},\ and\ \citenamefont
  {{Olinto}}}]{Jedamzik1}%
  \BibitemOpen
  \bibfield  {author} {\bibinfo {author} {\bibfnamefont {K.}~\bibnamefont
  {{Jedamzik}}}, \bibinfo {author} {\bibfnamefont {V.}~\bibnamefont
  {{Katalini{\'c}}}}, \ and\ \bibinfo {author} {\bibfnamefont {A.~V.}\
  \bibnamefont {{Olinto}}},\ }\href {\doibase 10.1103/PhysRevD.57.3264}
  {\bibfield  {journal} {\bibinfo  {journal} {\prd}\ }\textbf {\bibinfo
  {volume} {57}},\ \bibinfo {pages} {3264} (\bibinfo {year} {1998})},\ \Eprint
  {http://arxiv.org/abs/astro-ph/9606080} {astro-ph/9606080} \BibitemShut
  {NoStop}%
\bibitem [{\citenamefont {{Yamazaki}}\ and\ \citenamefont
  {{Kusakabe}}(2012)}]{yamazaki3}%
  \BibitemOpen
  \bibfield  {author} {\bibinfo {author} {\bibfnamefont {D.~G.}\ \bibnamefont
  {{Yamazaki}}}\ and\ \bibinfo {author} {\bibfnamefont {M.}~\bibnamefont
  {{Kusakabe}}},\ }\href {\doibase 10.1103/PhysRevD.86.123006} {\bibfield
  {journal} {\bibinfo  {journal} {\prd}\ }\textbf {\bibinfo {volume} {86}},\
  \bibinfo {eid} {123006} (\bibinfo {year} {2012})},\ \Eprint
  {http://arxiv.org/abs/1212.2968} {arXiv:1212.2968 [astro-ph.CO]} \BibitemShut
  {NoStop}%
\bibitem [{\citenamefont {{Yamazaki}}(2014)}]{yamazaki2}%
  \BibitemOpen
  \bibfield  {author} {\bibinfo {author} {\bibfnamefont {D.~G.}\ \bibnamefont
  {{Yamazaki}}},\ }\href {\doibase 10.1103/PhysRevD.89.083528} {\bibfield
  {journal} {\bibinfo  {journal} {\prd}\ }\textbf {\bibinfo {volume} {89}},\
  \bibinfo {eid} {083528} (\bibinfo {year} {2014})},\ \Eprint
  {http://arxiv.org/abs/1404.5310} {arXiv:1404.5310} \BibitemShut {NoStop}%
\bibitem [{\citenamefont {{Yamazaki}}\ \emph {et~al.}(2006)\citenamefont
  {{Yamazaki}}, \citenamefont {{Ichiki}}, \citenamefont {{Umezu}},\ and\
  \citenamefont {{Hanayama}}}]{yamazaki4}%
  \BibitemOpen
  \bibfield  {author} {\bibinfo {author} {\bibfnamefont {D.~G.}\ \bibnamefont
  {{Yamazaki}}}, \bibinfo {author} {\bibfnamefont {K.}~\bibnamefont
  {{Ichiki}}}, \bibinfo {author} {\bibfnamefont {K.-I.}\ \bibnamefont
  {{Umezu}}}, \ and\ \bibinfo {author} {\bibfnamefont {H.}~\bibnamefont
  {{Hanayama}}},\ }\href {\doibase 10.1103/PhysRevD.74.123518} {\bibfield
  {journal} {\bibinfo  {journal} {\prd}\ }\textbf {\bibinfo {volume} {74}},\
  \bibinfo {eid} {123518} (\bibinfo {year} {2006})},\ \Eprint
  {http://arxiv.org/abs/astro-ph/0611910} {astro-ph/0611910} \BibitemShut
  {NoStop}%
\bibitem [{\citenamefont {{Kim}}\ \emph {et~al.}(1996)\citenamefont {{Kim}},
  \citenamefont {{Olinto}},\ and\ \citenamefont {{Rosner}}}]{olinto}%
  \BibitemOpen
  \bibfield  {author} {\bibinfo {author} {\bibfnamefont {E.-J.}\ \bibnamefont
  {{Kim}}}, \bibinfo {author} {\bibfnamefont {A.~V.}\ \bibnamefont {{Olinto}}},
  \ and\ \bibinfo {author} {\bibfnamefont {R.}~\bibnamefont {{Rosner}}},\
  }\href {\doibase 10.1086/177667} {\bibfield  {journal} {\bibinfo  {journal}
  {\apj}\ }\textbf {\bibinfo {volume} {468}},\ \bibinfo {pages} {28} (\bibinfo
  {year} {1996})},\ \Eprint {http://arxiv.org/abs/astro-ph/9412070}
  {astro-ph/9412070} \BibitemShut {NoStop}%
\bibitem [{\citenamefont {{Kahniashvili}}\ \emph {et~al.}(2010)\citenamefont
  {{Kahniashvili}}, \citenamefont {{Tevzadze}}, \citenamefont {{Sethi}},
  \citenamefont {{Pandey}},\ and\ \citenamefont {{Ratra}}}]{tinaver1}%
  \BibitemOpen
  \bibfield  {author} {\bibinfo {author} {\bibfnamefont {T.}~\bibnamefont
  {{Kahniashvili}}}, \bibinfo {author} {\bibfnamefont {A.~G.}\ \bibnamefont
  {{Tevzadze}}}, \bibinfo {author} {\bibfnamefont {S.~K.}\ \bibnamefont
  {{Sethi}}}, \bibinfo {author} {\bibfnamefont {K.}~\bibnamefont {{Pandey}}}, \
  and\ \bibinfo {author} {\bibfnamefont {B.}~\bibnamefont {{Ratra}}},\ }\href
  {\doibase 10.1103/PhysRevD.82.083005} {\bibfield  {journal} {\bibinfo
  {journal} {\prd}\ }\textbf {\bibinfo {volume} {82}},\ \bibinfo {eid} {083005}
  (\bibinfo {year} {2010})},\ \Eprint {http://arxiv.org/abs/1009.2094}
  {arXiv:1009.2094 [astro-ph.CO]} \BibitemShut {NoStop}%
\bibitem [{\citenamefont {{Kahniashvili}}\ \emph {et~al.}(2011)\citenamefont
  {{Kahniashvili}}, \citenamefont {{Tevzadze}},\ and\ \citenamefont
  {{Ratra}}}]{tinaver2}%
  \BibitemOpen
  \bibfield  {author} {\bibinfo {author} {\bibfnamefont {T.}~\bibnamefont
  {{Kahniashvili}}}, \bibinfo {author} {\bibfnamefont {A.~G.}\ \bibnamefont
  {{Tevzadze}}}, \ and\ \bibinfo {author} {\bibfnamefont {B.}~\bibnamefont
  {{Ratra}}},\ }\href {\doibase 10.1088/0004-637X/726/2/78} {\bibfield
  {journal} {\bibinfo  {journal} {\apj}\ }\textbf {\bibinfo {volume} {726}},\
  \bibinfo {eid} {78} (\bibinfo {year} {2011})},\ \Eprint
  {http://arxiv.org/abs/0907.0197} {arXiv:0907.0197 [astro-ph.CO]} \BibitemShut
  {NoStop}%
\bibitem [{\citenamefont {{Hu}}\ and\ \citenamefont {{White}}(1997)}]{hu}%
  \BibitemOpen
  \bibfield  {author} {\bibinfo {author} {\bibfnamefont {W.}~\bibnamefont
  {{Hu}}}\ and\ \bibinfo {author} {\bibfnamefont {M.}~\bibnamefont {{White}}},\
  }\href {\doibase 10.1103/PhysRevD.56.596} {\bibfield  {journal} {\bibinfo
  {journal} {\prd}\ }\textbf {\bibinfo {volume} {56}},\ \bibinfo {pages} {596}
  (\bibinfo {year} {1997})},\ \Eprint {http://arxiv.org/abs/astro-ph/9702170}
  {astro-ph/9702170} \BibitemShut {NoStop}%
\bibitem [{\citenamefont {{Jedamzik}}\ \emph {et~al.}(2000)\citenamefont
  {{Jedamzik}}, \citenamefont {{Katalini{\'c}}},\ and\ \citenamefont
  {{Olinto}}}]{jedamzikolinto2}%
  \BibitemOpen
  \bibfield  {author} {\bibinfo {author} {\bibfnamefont {K.}~\bibnamefont
  {{Jedamzik}}}, \bibinfo {author} {\bibfnamefont {V.}~\bibnamefont
  {{Katalini{\'c}}}}, \ and\ \bibinfo {author} {\bibfnamefont {A.~V.}\
  \bibnamefont {{Olinto}}},\ }\href {\doibase 10.1103/PhysRevLett.85.700}
  {\bibfield  {journal} {\bibinfo  {journal} {Physical Review Letters}\
  }\textbf {\bibinfo {volume} {85}},\ \bibinfo {pages} {700} (\bibinfo {year}
  {2000})},\ \Eprint {http://arxiv.org/abs/astro-ph/9911100} {astro-ph/9911100}
  \BibitemShut {NoStop}%
\end{thebibliography}%
\end{document}